\documentclass[trackchanges,twocolumn,resetfootnote]{aastex701}
\DeclareRobustCommand{\ion}[2]{\textup{#1\,\textsc{\lowercase{#2}}}}
\newcommand*\element[1][]{%
  \def\aa@element@tr{#1}%
  \aa@element
}
\usepackage{amsmath}
\usepackage{booktabs} 
\usepackage{tablefootnote}
\usepackage{lipsum}
\usepackage{threeparttable}
\usepackage{xeCJK}
\usepackage{subcaption}
\usepackage{graphicx}

\begin{document}


\title{Resolved Dust in $z\approx1$ Galaxies with JWST/MIRI MRS:\\ Survey Description and First View on PAHs, Mid-IR Atomic Emission, and Warm H$_2$}

\correspondingauthor{Wuji Wang}
\email{wujiwang@ipac.caltech.edu}

\author[0000-0002-7964-6749]{Wuji Wang (王无忌)}
\affiliation{Caltech/IPAC, 1200 E. California Blvd. Pasadena, CA 91125, USA}
\email{}

\author[0000-0002-9382-9832]{Andreas L. Faisst}
\affiliation{Caltech/IPAC, 1200 E. California Blvd. Pasadena, CA 91125, USA}
\email{}

\author[0000-0001-8490-6632]{Thomas S.-Y. Lai (賴劭愉)}
\affiliation{Caltech/IPAC, 1200 E. California Blvd. Pasadena, CA 91125, USA}
\email{}

\author[0000-0002-4462-0709]{Kyle Finner}
\affiliation{Caltech/IPAC, 1200 E. California Blvd. Pasadena, CA 91125, USA}
\email{}

\author[0000-0002-6149-8178]{Jed McKinney}
\affiliation{Department of Astronomy, The University of Texas at Austin, Austin, TX, USA}
\affiliation{Cosmic Frontier Center, The University of Texas at Austin, Austin, TX 78712, USA}
\email{}

\author[0000-0003-4569-2285]{Andrew J. Battisti}
\affiliation{International Centre for Radio Astronomy Research, University of Western Australia, 7 Fairway, Crawley, WA 6009, Australia}
\affiliation{Research School of Astronomy and Astrophysics, Australian National University, Cotter Road, Weston Creek, ACT 2611, Australia}
\email{}

\author[0000-0003-4702-7561]{Irene Shivaei}
\affiliation{Centro de Astrobiolog\'ia (CAB), CSIC-INTA, Carretera de Ajalvir km 4, Torrej\'on de Ardoz, E-28850, Madrid, Spain}
\email{}

\author[0000-0002-3952-8588]{Leindert A. Boogaard}
\affiliation{Leiden Observatory, Leiden University, PO Box 9513, NL-2300 RA Leiden, The Netherlands}
\email{}

\author[0000-0001-9773-7479]{Daizhong Liu}
\affiliation{Purple Mountain Observatory, Chinese Academy of Sciences, Nanjing, China}
\email{}

\author[0000-0001-8792-3091]{Yu-Heng Lin}
\affiliation{Caltech/IPAC, 1200 E. California Blvd. Pasadena, CA 91125, USA}
\email{ianlin@ipac.caltech.edu}

\author[0009-0004-1270-2373]{Lun-Jun Liu}
\email{lliu@caltech.edu}
\affiliation{California Institute of Technology, 1200 E. California Blvd., Pasadena, CA, 91125 USA}

\author[0000-0002-4872-2294]{Georgios Magdis}
\affiliation{Cosmic Dawn Center (DAWN), Copenhagen, Denmark}
\affiliation{Department for Space Research and Space Technology (DTU Space), Technical University of Denmark, Elektrovej 327, DK-2800 Kgs. Lyngby, Denmark}
\affiliation{Niels Bohr Institute, University of Copenhagen, Jagtvej 128, DK-2200 Copenhagen, Denmark}
\email{}

\author[0000-0002-1047-9583]{Vincenzo Mainieri}
\affiliation{European Southern Observatory, Karl-Schwarzschild-Strasse 2, 85748 Garching bei M\"unchen, Germany}
\email{}

\author[0000-0002-8813-9116]{Alexander Rodriguez}
\affiliation{Department of Astronomy, University of Michigan, Ann Arbor, MI, USA}
\email{}

\author[0000-0002-0000-6977]{John D. Silverman}
\email{}
\affiliation{Kavli Institute for the Physics and Mathematics of the Universe (Kavli IPMU, WPI), UTIAS, Tokyo Institutes for Advanced Study, University of Tokyo, Chiba, 277-8583, Japan}
\affiliation{Department of Astronomy, Graduate School of Science, The University of Tokyo, 7-3-1 Hongo, Bunkyo, Tokyo 113-0033, Japan}
\affiliation{Center for Data-Driven Discovery, Kavli IPMU (WPI), UTIAS, The University of Tokyo, Kashiwa, Chiba 277-8583, Japan}
\affiliation{Center for Astrophysical Sciences, Department of Physics \& Astronomy, Johns Hopkins University, Baltimore, MD 21218, USA}

\author[0000-0002-9252-114X]{Zhaoxuan Liu}
\email{}
\affiliation{Kavli Institute for the Physics and Mathematics of the Universe (Kavli IPMU, WPI), UTIAS, Tokyo Institutes for Advanced Study, University of Tokyo, Chiba, 277-8583, Japan}
\affiliation{Department of Astronomy, Graduate School of Science, The University of Tokyo, 7-3-1 Hongo, Bunkyo, Tokyo 113-0033, Japan}
\affiliation{Center for Data-Driven Discovery, Kavli IPMU (WPI), UTIAS, The University of Tokyo, Kashiwa, Chiba 277-8583, Japan}
\affiliation{Universit\'e Paris-Saclay, Universit\'e Paris Cité, CEA, CNRS, AIM, 91191, Gif-sur-Yvette, France}

\author[0000-0002-2419-3068]{Minju M. Lee}
\email{}
\affiliation{Cosmic Dawn Center (DAWN), Copenhagen, Denmark}
\affiliation{Department for Space Research and Space Technology (DTU Space), Technical University of Denmark, Elektrovej 327, DK-2800 Kgs. Lyngby, Denmark}

\author[0000-0002-0236-919X]{Ghassem Gozaliasl}
\email{}
\affiliation{Department of Computer Science, Aalto University, PO Box 15400,
Espoo, FI-00076, Finland}
\affiliation{Department of Physics, University of Helsinki, P. O. Box 64, FI00014 Helsinki, Finland}

\author[0000-0002-3560-8599]{Maximilien Franco}
\affiliation{Universit\'e Paris-Saclay, Universit\'e Paris Cité, CEA, CNRS, AIM, 91191, Gif-sur-Yvette, France}
\email{}


\begin{abstract}
Dust is a key component of galaxies that regulates their thermal balance and, consequently, star formation and the build-up of stellar mass. Polycyclic aromatic hydrocarbons (PAHs) are responsible for reprocessing radiative energy of the galaxies thus tracing dust evolution. Using JWST/MIRI MRS, we present the first resolved view of the PAHs and mid-infrared emission lines in a sample of eight $z\approx1$ galaxies. A key novelty is our ability to directly map PAH emission near the end of cosmic noon at JWST's limits. The sample is selected to be on the star-forming main-sequence with stellar masses $M_{\star}=10^{10.6-11.2}\,M_{\odot}$ and infrared luminosities $L_{\rm IR}=10^{11.5-11.9}\,L_{\odot}$. Two of them contain active galactic nuclei (AGN), and two are interacting systems. We detect and quantify primary PAH emissions from 3.3$\,\rm \mu m$ to 11.3$\,\rm\mu m$ throughout the galaxies, alongside atomic fine structure lines (Ar, Ne, and Fe),  Br$\alpha$, and H$_{2}$ rotational transitions. Through PAH ratio diagnostics and comparison to theoretical models, we qualitatively probe the physical properties of PAH molecules, i.e., size and charge. The AGN and mergers in our sample exhibit a higher fraction of neutral PAHs, possibly related to high radiation intensity and/or shocks, as suggested by increased atomic and H$_{2}$ line ratios. Leveraging the IFU data, we find that the grain sizes of the centrally located PAHs tend to be larger and less ionized than those in the outskirts of the galaxies. Finally, we compare our results to observations of PAHs in local and similar-redshift galaxies, revealing a potential evolutionary trend when controlling $L_{\rm IR}$.

\end{abstract}



\section{Introduction}\label{sec:intro} 

\begin{table*}
 \caption{Properties of the $z\approx1$ galaxies analyzed in this paper.}\label{tab:sample}
 \centering
\begin{tabular}{c c c c c c c }
\hline
\hline
ID  & RA, DEC (J2000)      & Redshift & $\log(M_{*}/M_{\odot})$ & SFR$_{\rm UV+IR}$ & $\log (L_{\rm UV}/L_{\odot})$ & $\log (L_{\rm 8-1000\mu m}/L_{\odot})$   \\
    & (hh:mm:sss, dd:mm:ss) & $z_{\rm [\ion{Ar}{ii}]}$ &   & $M_{\odot}\,\rm yr^{-1}$ &  &   \\
(1) & (2) & (3) & (4) & (5) & (6) & (7) \\

\hline
COSMOS 1018$^{\rm A}$ &                           & & $11.12\pm0.08$ & 86 & 10.04  & 11.92  \\
A           & 10:00:31.162, 02:06:39.19 & 0.9365 &       & & &  \\
B           & 10:00:31.130, 02:06:38.39 & 0.9375 &       & & & \\
\cline{1-1}
COSMOS 1295 &                           & & $11.08\pm0.02$ & 47 & 10.17  & 11.64  \\
A & 09:59:26.405, 02:11:25.01           & 0.9473 &       & & &  \\
B & 09:59:26.458, 02:11:25.27           & 0.9493 &       & & & \\
\cline{1-1}
COSMOS 1148$^{\rm A}$ & 10:00:04.530, 02:08:52.14 & 0.9584 & $11.16\pm0.12$ & 42  & 10.53 & 11.55   \\
COSMOS 1025 & 10:00:27.558, 02:06:50.20 & 0.9355 & $10.67\pm0.02$ & 39  & 10.13 & 11.56  \\
COSMOS 996  & 10:00:48.940, 02:06:03.28 & 0.9307 & $11.01\pm0.03$ & 76  & 10.45 & 11.85   \\
COSMOS 844  & 09:59:45.304, 02:02:07.77 & 0.9381 & $10.81\pm0.02$ & 61  & 9.45  & 11.78  \\
COSMOS 1757 & 10:00:22.702, 02:20:10.79 & 0.9290 & $10.76\pm0.02$ & 53  & 9.39  & 11.72  \\
COSMOS 1346 & 10:00:35.563, 02:11:50.91 & 0.9259 & $11.18\pm0.02$ & 33  & 9.94  & 11.49  \\

 \hline
\end{tabular}
\parbox{\textwidth}{
\smallskip
\textbf{Notes:} 
(1) ID of the galaxy in our sample. AGN are labeled with a subscript ``A''.
(2) Coordinate (incl. merger components).
(3) Systemic redshifts based on [\ion{Ar}{ii}] (typical uncertainty is 0.0001).
(4) Stellar masses SED fitting with \texttt{CIGALE} using MIRI photometry from this work.
(5) Total SFR based on UV$+$IR luminosity.
(6-7) UV (at $1600\rm \,\AA$) and total infrared luminosity ($8-1000\,\mu$m) derived from the fitted SED. } 
\end{table*}

Dust plays a crucial role in the formation and evolution of galaxies by providing the grounds for converting molecular gas into stars \citep[][]{Galliano_2018}. The UV radiation in the interstellar medium (ISM) is absorbed by dust and re-emitted in the infrared. This emission dominates the galaxy luminosity at longer wavelengths, especially in star-forming galaxies \citep[][]{Calzetti_2000, Salim_2022}. Produced by AGB stars and supernovae alongside metals in mature galaxies, dust is a key probe of the chemical evolution of the ISM \citep[e.g.,][]{Ferrarotti_2006}. In addition, the growth of dust in the ISM through accretion of atoms and molecules is also found to be an important path for dust growth \citep[e.g.][]{Zhukovska_2008,Galliano_2018}.

Polycyclic Aromatic hydrocarbons (PAHs) are a special species of dust molecules typically composed of several tens to thousands of carbon atoms \citep[sizes of $\lesssim 5\times 10^{-3}\,\rm \mu m$, see, e.g., the reviews by][]{LiAigen_2020,Tielens_2008,Tielens_2026}\footnote{Sometimes PAHs are referred to as ``nano-grains''.}.
Despite the low mass fraction ($\lesssim5\%$) of PAHs to the total dust budget, their luminosities account up to $20\%$ of the total infrared emissions, making PAHs a dominating source of heating in the ISM \citep[][]{Helou_2001,Smith_2007,Lai_2020,Shivaei_2024}. The various stretching and bending modes of C-C and C-H molecules in PAHs produce unique emission signatures in the mid-IR between 3.3 to 17$ \rm \mu m$. The relative strength of the PAH emission bands are a useful diagnostic tool for studying grain properties, for example, size and charges of these PAH molecules \citep[e.g.][]{Draine_2007,Draine_2021,Maragkoudakis_2020}. These intrinsic PAH properties are related to the broader ISM environment, for example, radiation intensity and metallicity, which controls the formation and processing of the dust. Shattering of larger dust grains, $\sim0.1\rm \, \mu m$, is found to be a prominent path for the formation of $\sim10^{-3}\rm \, \mu m$ PAH molecules \citep[e.g., due to supernova shocks,][]{Jones_1996,Wiebe_2014,Asano_2013}. In low metallicity environments, in-situ grain growth in dense media is a vital pathway for PAH formation and survival \citep[][]{Whitcomb_2024,Lai_2025,Tarantino_2026}. The observed grain populations result from the equilibrium between the destruction and formation of PAHs controlled by these mechanisms. PAHs are also entangled dynamically with the evolution of large non-aromatic dust grains \citep[e.g.][]{Rau_2019}.

The close relations between PAHs, star formation, and the radiation properties of the ISM makes them a useful tool for understanding the buildup and evolution of the prototypical galaxies \citep[][]{Shipley_2016}. While significant efforts have been undertaken to study the PAH properties in more local galaxies, it is of particular interest to trace these aromatic molecules in an earlier stage of cosmic history. Redshifts $z\approx1-2$ mark the period of cosmic noon where the cosmic star formation density is at its peak \citep[][]{Madau_2014}.
Spatially unresolved PAH studies at higher redshifts have been performed by Spitzer in the pre-JWST era with samples limited to (ultra)luminous infrared galaxies, (U)LIRGs \citep[$L_{\rm IR}\gtrsim10^{11-12}\,L_{\odot}$;][]{Lutz_2005,Yan_2007,Rawlings_2013}.
Thanks to its sensitivity at mid-IR wavelengths, JWST pushes the limit of PAHs studies to the early universe.  For example, \citet{Shivaei_2024} investigated the relation between dust and galaxy properties at $0.7<z<2$ making use of the $5-25~\rm\mu m$ photometric band passes of JWST. \citet{Veilleux_2025} reported PAHs on circumgalactic scales at $z\sim0.5$. \citet{Lyu_2025} characterized the PAH3.3$\,\rm \mu m$ and aliphatic feature at 3.4$\,\rm \mu m$ out to $z<0.45$ with slitless spectroscopy. \citet{McKinney_2026} built a spectroscopic sample of PAHs in $0.6<z<2.5$ (U)LIRGs using the low resolution spectroscopic mode of MIRI.  

Despite the previous results, over the four years of operation, the spatially resolved view of PAHs using JWST IFU spectroscopy is limited to a few systems beyond local. For example, \citet{Young_2023} and \citet{Sajina_2025} targeted a dusty galaxy at $z=0.54$ where gas, dust, and star formation activities have been mapped. The PAHSPEC program observed five $z\sim1.1$ galaxies and found evidence of evolution in ISM dust properties compared to local \citep[][]{Donnan_2026,Lofaro_2026}. \citet{Spilker_2023} resolved the PAH3.3\,$\rm \mu m$ in a lensed galaxy at $z=4.2$ which provides evidence of different dust formation in the early universe. To systematically understand the interplay between star formation, AGN, and dust evolution, resolved observations on galactic scales are crucial \citep[][]{Lai_2022,Lai_2023,GarciaBernete_2024c}. In addition, galaxy mergers also influence the PAH demographic (e.g., size distribution and abundances), an effect that was inaccessible beyond the local universe prior to JWST \citep[][]{Murata_2017}.

This is the inaugural paper in a series to present the JWST/MIRI Medium Resolution Spectroscopy (MRS) study of spatially resolved PAHs and mid-IR spectroscopic features in a sample of eight prototypical star-forming main-sequence galaxies at the end of cosmic noon ($z\approx1$). In this paper, we describe the sample, observations, and data processing in Sect.~\ref{sec:sample_obs_data}. Sect.~\ref{sec:analysis} presents our spectral analysis of the line emissions and PAHs and the results are shown in Sect.~\ref{sec:results}. Sect.~\ref{sec:discuss} discusses the resolved PAH properties linked to the host galaxies for our sample and discusses the connection to other samples in the literature. Finally, we summarize and conclude in Sect.~\ref{sec:conclu}.

Throughout this work, we assume a flat $\Lambda$CDM cosmology with $H_{0} = 70\, \rm{km\,s^{-1}\,Mpc^{-1}}$ and $\Omega_{m}=0.3$. Following this cosmology, we will have $\rm{1\,arcsec\approx7.75\, kpc}$ at the redshifts ($0.92<z<0.96$) of this work. Stellar masses ($M_{\star}$) and star formation rates (SFR) are normalized to a \citet{chabrier_2003} IMF.

\section{Sample, Observations, and Data Processing} \label{sec:sample_obs_data}

\subsection{Sample Selection}\label{subsec:sample}

Our parent sample was selected from the JWST COSMOS-Web survey area \citep[GO-1727; PIs: Kartaltepe \& Casey;][]{Casey_2023} which provides JWST/NIRCam and MIRI imaging \citep[F115W, F150W, F277W, F444W, F770W;][]{Harish_2025,Franco_2026}. COSMOS-Web is part of the COSMOS field with rich multi-wavelength ancillary data \citep{Scoville_2007,koekemoer_2007,Ilbert_2013,Laigle_2016,Weaver_2022,Shuntov_2025}.
To ensure that all major PAH ($3.3$ to $11.3\,{\rm \mu m}$) and fine structure lines fall near the MIRI/MRS channel center, we further limited the redshift range to $0.92<z<0.98$. The spectroscopic redshifts, $z_{\rm spec}$, were obtained from the public COSMOS spectroscopy catalog by \citet{Khostovan_2026} \citep[including][]{George_2011,Hu_2021,Horowitz_2022,Saglia_2022}.
The galaxies were chosen to yield a good signal-to-noise ratio across the mid-IR range in a reasonable observation time with JWST.
This leads to a narrower sample containing star-forming galaxies with stellar masses $10^{11}\,M_{\odot}$, at the high-mass end of the $\log({\rm SFR})-\log(M_{\star})$ main-sequence relation at $z\approx1$ \citep[e.g.,][]{Speagle_2014,Schreiber_2015}.
For an optimal view of the galactic structures, we further selected the galaxies to be nearly face-on. 
The final sample includes eight galaxies between $0.92 < z < 0.96$.
Fig. \ref{fig:MS_sample} shows the parent and our selected sample on the $M_{\star}-$SFR plane and Table~\ref{tab:sample} summarizes the target sample properties.
The SFR and $M_{\star}$ reported in this paper are updated, compared to previous values from \citet{Weaver_2022}, by making use of the new MIRI data from this program \citep[Sect. \ref{subsec:sed_fit};][]{Faisst_2026_UVbump}.
Two galaxies were visually identified as merging systems (COSMOS~1018 and COSMOS~1295, Sect. \ref{subsec:morphology}). COSMOS~1148 was identified as a possible AGN due to a {\it Chandra} X-ray detection. 
As discussed in Sect. \ref{subsec:res_AGN}, we identify another possible AGN (COSMOS~1018) based on the MIRI/MRS spectrum and MIRI photometry.

\begin{figure}[t!]
    \centering
    \includegraphics[width=\columnwidth,clip]{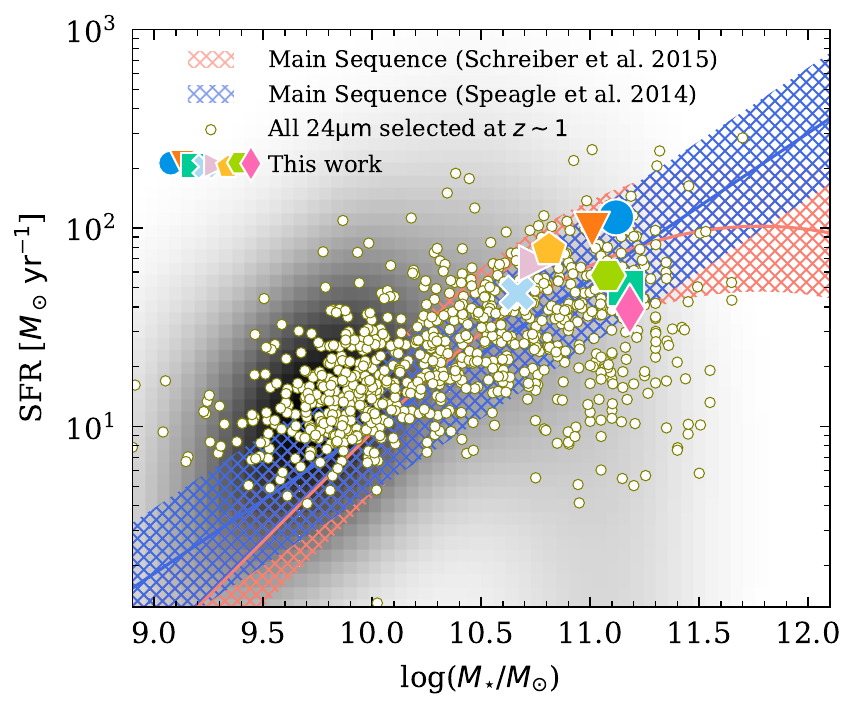}
    \caption{
    Sample studied in this work (large colored symbols, same used throughout this work) compared to the parent sample (Spitzer/MIPS $24\,\rm \mu m$ detections; white circles) on the star-forming main-sequence (from \citealt{Schreiber_2015} and \citealt{Speagle_2014} for comparison). The gray background includes all spectroscopically confirmed COSMOS galaxies at $z=0.9-1.0$. 
    }
\label{fig:MS_sample}
\end{figure}

\begin{figure*}[ht!]
    \centering
    \includegraphics[width=0.9\textwidth,clip]{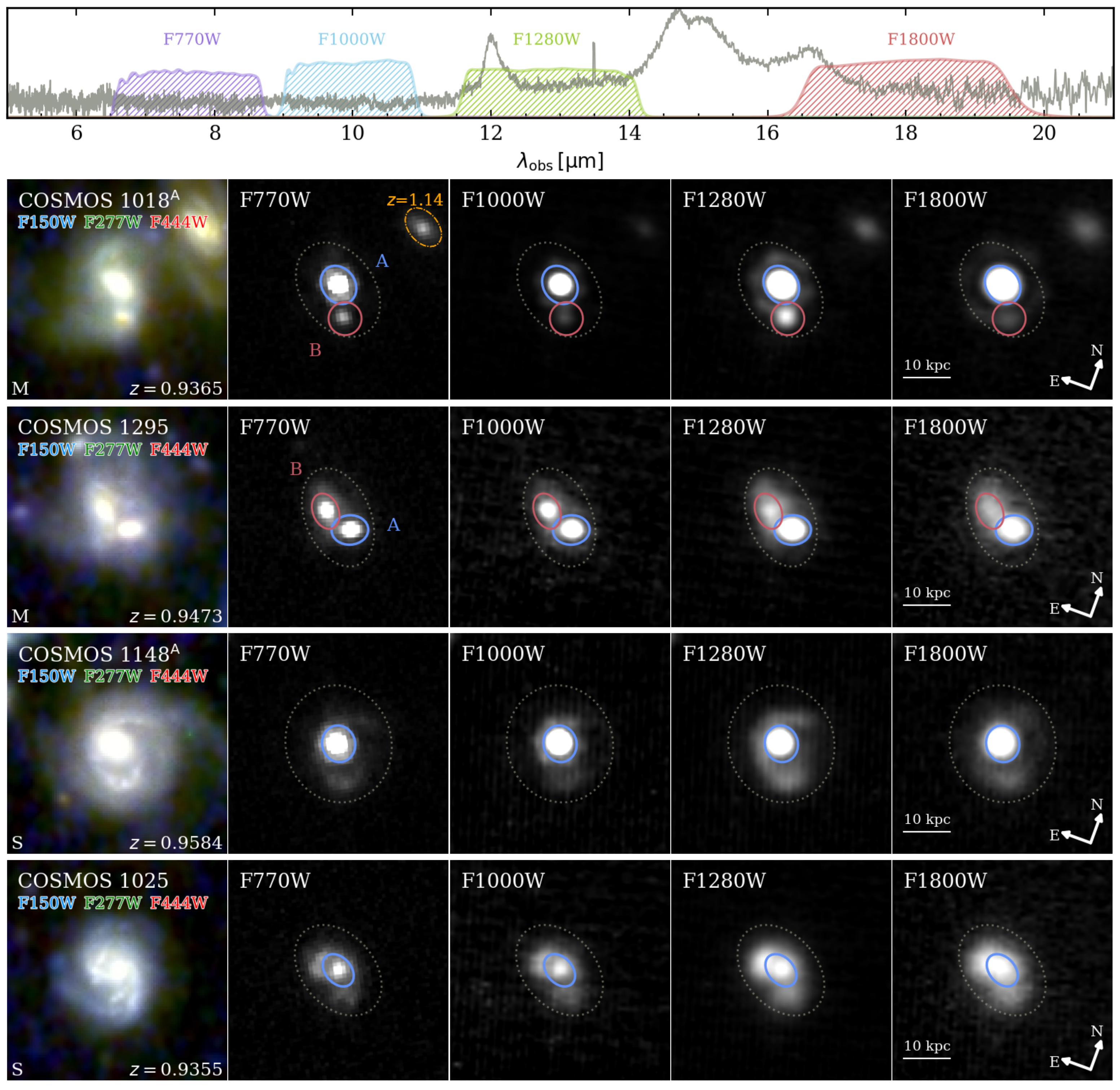}
    \caption{
    {\em Top panel:} Transmission of MIRI filters overlaid on the spectrum of COSMOS~996 as an example.
    {\em Bottom panels:} JWST imaging of galaxies in our sample. Each column from left to right shows the composite RGB image from NIRCam F150W$+$F277W$+$F444W (taken from COSMOS-Web), and the individual MIRI filters (F770W, F1000W, F1280W, and F1800W from this program). Each cutout is $6\arcsec\times6\arcsec$.
    Mergers are indicated with ``M'', non-mergers are indicated with ``D'' or ``S`` (in case of double peaked argon line). AGN's are marked with ``A''.
    The blue ellipses mark the ``center'' extraction apertures (red ellipses are added for mergers for the second component). Dashed ellipses show the extraction regions of the ``outskirt'' regions.
    }
    \label{fig:nircam_rgb}
    \addtocounter{figure}{-1}
\end{figure*}

\subsection{Observations and Data Processing}\label{subsec:data_process}

The eight targets were observed by JWST/MIRI MRS in the Cycle 3 General Observer program GO-4761 (PI: A. Faisst) between UT 2025 April 05 and May 08. All three sub-bands were used to cover the observed wavelength range of $5-28\,\mu$m (rest-frame $\sim2.6-14\,\mu$m) with the same exposure time per sub-band (ranging from $0.5$ to $1.4~$hr, depending on target).
We adopted a 4-point dither pattern optimized for extended sources for the science exposures.
Offset simultaneous MIRI imaging in F1000W was obtained in parallel with the MRS science observations for all three sub-band exposures (totaling to $1.5-4.2~$hr per pointing -- adding to the COSMOS legacy value).
We assigned dedicated MRS background observations to each target using a 2-point dither pattern ({\it i.e.} half of the exposure time as the science target), which we found sufficient for background subtraction of the main MIRI MRS exposures. The location and position angle of the MRS background observations were chosen to cover the science targets with additional MIRI imaging in different pass bands. Specifically, we added F1000W, F1280W, and F1800W to the existing COSMOS-Web F770W imaging for our targets to enable more robust high-resolution photometric constraints of the mid-IR continuum, specifically the silicate absorption. For COSMOS-1757, we used F1500W instead of F1800W, as the latter was already available from the COSMOS/PRIMER program with shallower exposure \citep[GO 1837; PI: J. Dunlop;][]{Donnan_2024}. By construction, for each of these filters, the exposure time is the same as for the MIRI MRS sub-bands.

The JWST/MIRI MRS data are downloaded from the Mikulski Archive for Space Telescope (MAST): \dataset[https://doi.org/10.17909/9nf0-je75]{https://doi.org/10.17909/9nf0-je75} and processed using the JWST pipeline version 1.19.1 with reference file \texttt{jwst\_1410.pmap}.
We followed the reduction procedure from \citet{Law_2025_reduction}. The dedicated MIRI MRS backgrounds were subtracted using a pixel-based background subtraction method.
The long-exposure MIRI MRS observations are known to be impacted by cosmic rays. The artifacts from unflagged cosmic rays by the reduction pipeline will leave stripe patterns in the horizontal direction along the detector. We removed these stripes following the method described in \citet{Spilker_2023} by fitting and subtracting the stripe patterns at each wavelength channel with the sources masked.
Finally, we generated MIRI MRS data cubes for each galaxy with \texttt{cube\_build.output\_type = 'multi'} at Stage 3, which outputs one ``uber cube'' where all four channels are resampled to the same pixel scale. This setup facilitates the spatial mapping. We also examined the spectra from cubes with different constructions (\texttt{cube\_build.output\_type = 'band'} and \texttt{'channel'}), but we found that the differences are minimal.
In the later analysis, we accounted for the wavelength-depended point spread function (PSF) over $23~\mu$m ($\sim0.3-0.9\arcsec$ FWHM) by following the recipe of \citet{Lai_2022} and performed PSF matching to the PSF at 22.1~$\mu$m, which is equivalent to the observed wavelength of the PAH 11.3~$\mu$m feature. 
For astrometry quantification and correction, we performed a two-step procedure:
{\it (i)} we aligned MIRI images obtained by the dedicated background exposures using GAIA \citep{Gaia_DR3} stars;
{\it (ii)} we aligned the corrected MIRI images to the MRS cube by collapsing it in $\lambda$-direction over the respective filter transmission curve and by comparing the centroids of stars. The median offsets for our imaging and MRS IFU observations before the correction are $\rm \sim200~mas$. With our astrometry correction, the offsets are reduced to $\rm \sim28~mas$, which is the final astrometry precision of our images and MRS cubes.

\begin{figure*}[t!]
    \centering
    \includegraphics[width=0.9\textwidth,clip]{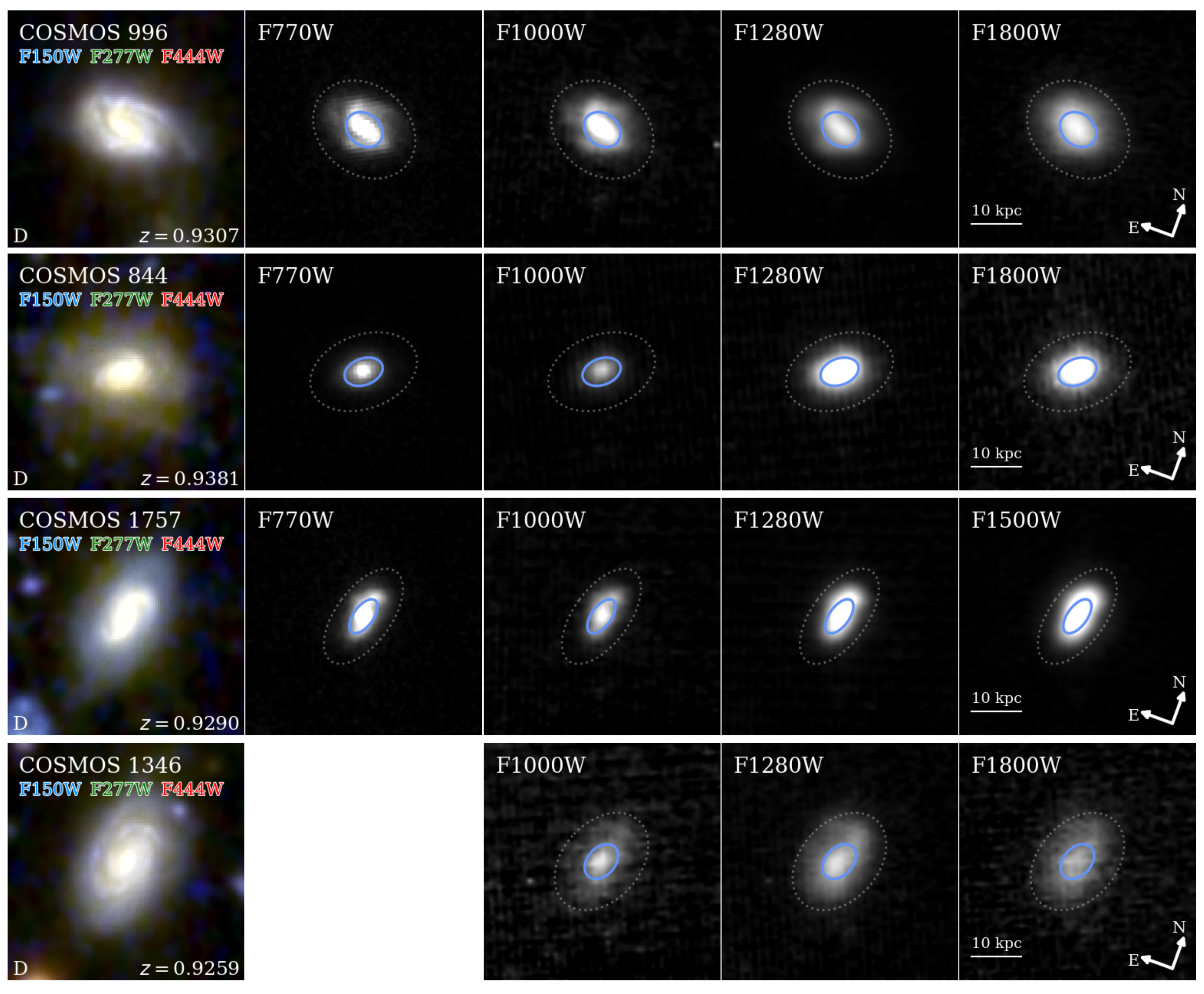}
    \caption{Continued. For COSMOS~1757, we obtained F1500W imaging as F1800W is available from PRIMER/COSMOS. COSMOS~1346 falls on the detector edge in F770W.
    }
\end{figure*}

We further check the absolute and relative flux calibrations by comparing the photometry from the data cubes and the MIRI imaging.
To this end, we create synthetic images from the cubes by convolving with the respective filter curves and compare the fluxes measured in apertures from these synthetic images ($f_{\rm cube}$) and the corresponding MIRI images ($f_{\rm ima}$).
We find a consistency $|f_{\rm cube}-f_{\rm ima}|/f_{\rm ima}\approx14\%$ on average. This difference could be accounted for by the different image and spectral PSFs.

\subsection{Stellar Masses and SFRs}\label{subsec:sed_fit}

Leveraging the new MIRI imaging photometry data, we used \texttt{CIGALE} \citep[][]{Boquien_2019,Burgarella2025} to fit the spectral energy distributions (SEDs) of our sample. The details of the SED fitting are described in \citet[][]{Faisst_2026_UVbump}.
In brief, the UV to far-IR photometry data are based on COSMOS2020 \citep{Weaver_2022} and the ``super-deblended'' \citep[][]{Jin_2018} catalogs. The latter contain IR and sub-mm observations, including 850$\,\mu$m JCMT/SCUBA-2 data, which cover the longest wavelength regions for our targets and constrain the cold dust emission peak.
Additionally, we included the MIRI F770W photometry from COSMOS-Web, F1800W from COSMOS/PRIMER for target COSMOS~1757, and F1000W, F1280W, and F1800W from this program for the remaining targets.
The MIRI fluxes from this program were extracted from the gray-dotted apertures (Fig. \ref{fig:nircam_rgb} and Sect. \ref{subsec:aper_ana}) and are listed in Table~\ref{tab:ima_photo}.
\texttt{CIGALE} is then run for a large grid of model parameters assuming a delayed star formation history with various stellar population ages, dust temperatures and attenuations, metallicities, and mid-IR AGN contributions.

We note that SED-derived SFRs depend on the assumed star formation (history) models \citep[see][]{Faisst_2026_UVbump}. In this work, to compare to the literature, we therefore only focus on the combined UV$+$IR total SFRs (SFR$_{\rm UV+IR}$), which are derived from the sum of the UV luminosity \citep[$1600\,{\rm \AA}$;][]{Kennicutt_1998} and the total infrared luminosity integrated from $8-1000\,{\rm \AA}$ ($L_{\rm IR}$, converted to SFR using the calibration by \citealt{Inami_2022}).
Table \ref{tab:sample} summarizes the stellar mass ($M_{\star}$) measurements from the \texttt{CIGALE} SED fit as well as total UV$+$IR SFRs for our galaxies.


\begin{table*}[t!]
\caption{Photometry from MIRI imaging for our sample. Fluxes are in $\rm \mu Jy$.}\label{tab:ima_photo}
\centering
\begin{tabular}{lccccc}
\hline
ID & F770W & F1000W & F1280W & F1500W & F1800W \\
\hline
COSMOS 1018 & 91.5 $\pm$ 0.8 & 156.7 $\pm$ 2.3 & 359.7 $\pm$ 0.5 & -- & 556.6 $\pm$ 12.2 \\
COSMOS 1295 & 30.5 $\pm$ 0.2 & 31.9 $\pm$ 2.0 & 133.5 $\pm$ 1.1 & -- & 126.6 $\pm$ 7.8 \\
COSMOS 1148 & 66.6 $\pm$ 0.6 & 80.0 $\pm$ 3.2 & 227.4 $\pm$ 4.9 & -- & 294.2 $\pm$ 8.6 \\
COSMOS 1025 & 29.9 $\pm$ 0.5 & 33.8 $\pm$ 3.2 & 215.0 $\pm$ 2.2 & -- & 189.4 $\pm$ 7.4 \\
COSMOS 996 & 54.5 $\pm$ 0.3 & 64.3 $\pm$ 1.8 & 409.2 $\pm$ 3.3 & -- & 339.7 $\pm$ 10.5 \\
COSMOS 844 & 32.3 $\pm$ 0.4 & 32.4 $\pm$ 0.7 & 165.6 $\pm$ 2.6 & -- & 146.5 $\pm$ 5.4 \\
COSMOS 1757 & 33.9 $\pm$ 0.3 & 35.6 $\pm$ 1.6 & 190.9 $\pm$ 2.1 & 391.1 $\pm$ 1.3 & 152.7 $\pm$ 2.4 \\
COSMOS 1346 & -- & 29.1 $\pm$ 3.5 & 135.3 $\pm$ 3.8 & -- & 115.2 $\pm$ 5.8 \\

\hline
\end{tabular}
\end{table*}

\subsection{Identification of Mergers}\label{subsec:morphology}

Based on visual examination of the morphology from NIRCam imaging, we identify two mergers in our sample: COSMOS~1295 (major merger of roughly equal luminosity) and COSMOS~1018 (likely a minor merger). Further examination of the mergers from profile decomposition and resolved SED fitting will be presented in a future study.
We mark the mergers as ``M'' in Fig. \ref{fig:nircam_rgb}. Additionally we use ``D'' (or ``S'') for non-mergers to indicate whether the atomic fine structure lines (such as [\ion{Ar}{ii}]) are {\em double} or {\em single} peaked (Sect. \ref{subsec:z_sys_linefit}).

\subsection{The Environment near COSMOS 1018}\label{subsec:dens_1018}

We note that COSMOS 1018 could be in an over-dense region identified by a preliminary examination of the photometric redshifts using the COSMOS-Web viewer\footnote{\url{https://cosmos2025.iap.fr/fitsmap.html}}. Specifically, approximately $\gtrsim10$ galaxies are found to have $z_{\rm photo}\sim0.91-0.96$ in physical proximity of $\sim1\arcmin$ of COSMOS~1018.
In addition, we report a galaxy $\sim 2.9\arcsec$ away from COSMOS~1018 captured in the field of view (FOV) of our MIRI MRS channel 2, 3, and 4. In Fig. \ref{fig:nircam_rgb} we mark the position of that galaxy by an ellipse with size of $1.2\arcsec \times 0.8 \arcsec$. The integrated MIRI/MRS spectrum (see Appendix~\ref{app:spec_presen}) measures a spectroscopic redshift of $z=1.141$. Although this galaxy is in close angular projection it is not physically associated with the overdensity around COSMOS~1018. 

\begin{figure*}[t!]
    \centering
    \includegraphics[width=\textwidth,clip]{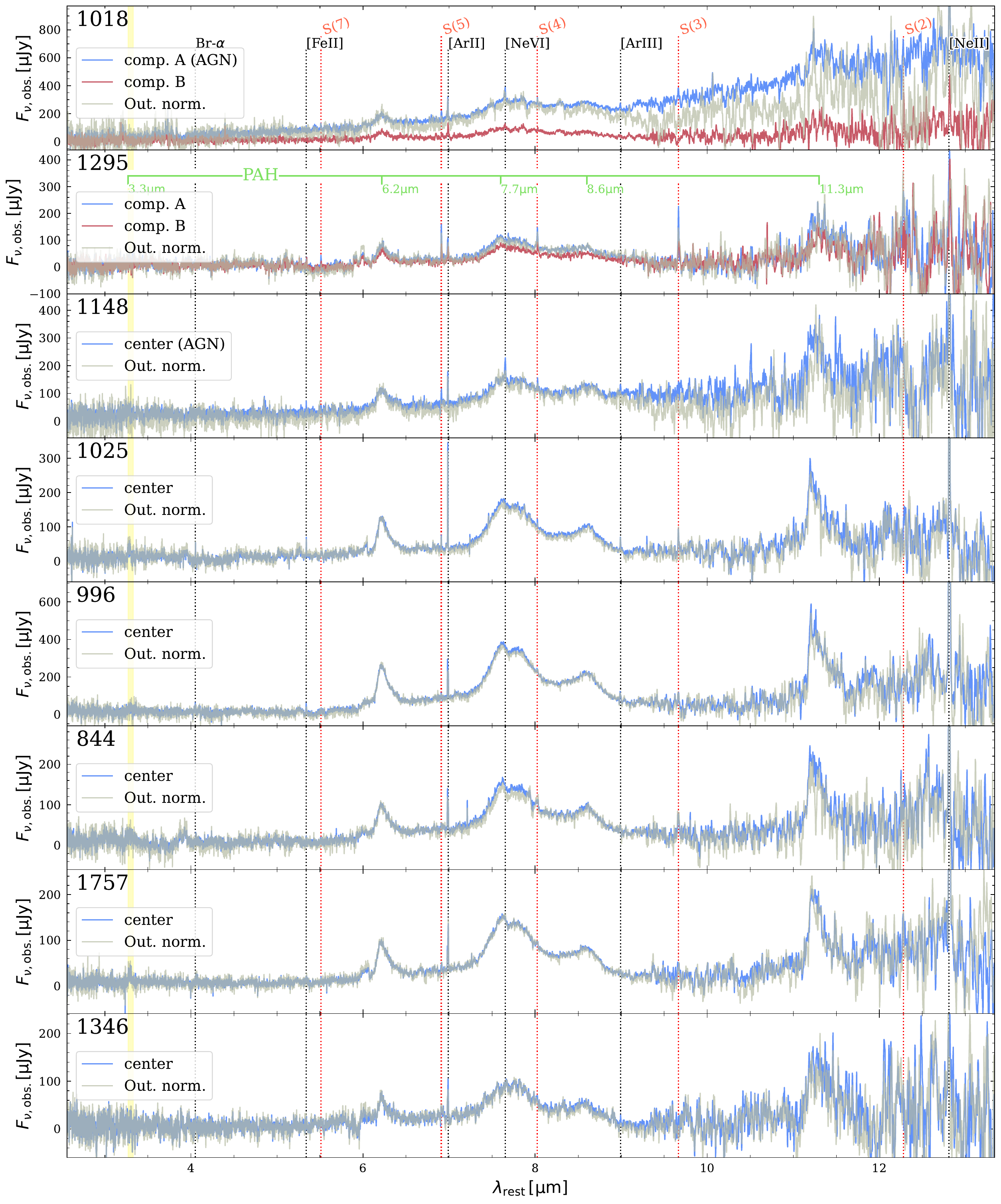}
    \caption{
    JWST/MIRI MRS spectra of our sample split in {\em central} (blue) and {\em outskirt} (gray) spatial regions. For the mergers, we have separated the two merging components. The spectra are normalized at $6.2\,{\rm \mu m}$. Various PAH bands, fine-structure line, and H$_2$ rotational lines are marked.
    \label{fig:spc_present}
    }
\end{figure*}

\begin{figure}
    \centering
    \includegraphics[width=\columnwidth,clip]{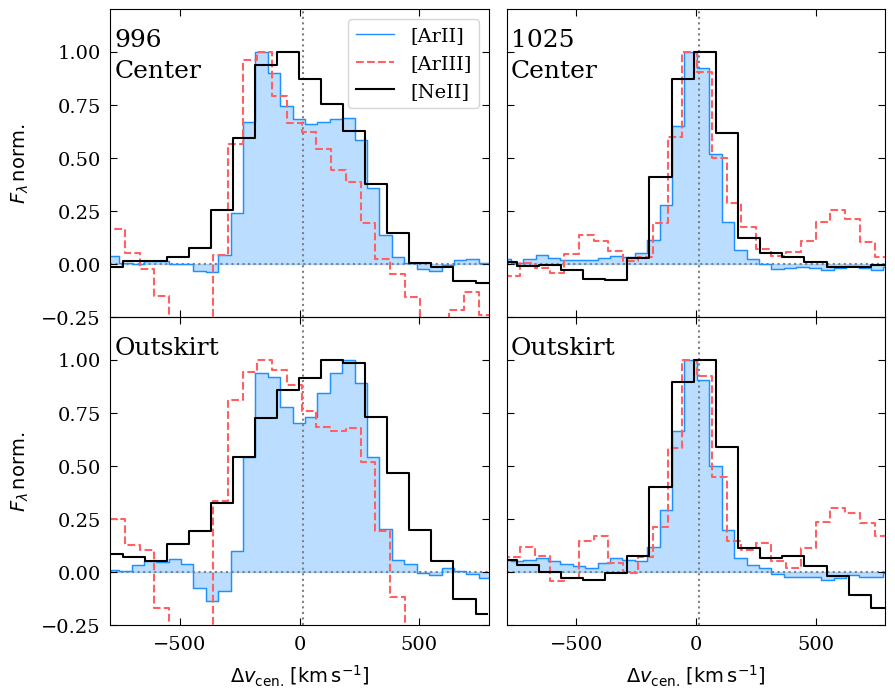}
    \caption{
    Fine-structure lines of two example non-mergers with (COSMOS~996; left) and without (COSMOS~1025; right) double-peaked \ion{Ar}{ii} lines. The central region and outskirts are shown. The spectra are continuum-subtracted locally and normalized to the line peak flux density. The $\Delta v=0\, \rm km\, s^{-1}$ marks the $z_{\rm sys}$ at the center. Figures for the remaining non-mergers in our sample can be found in Appendix~\ref{app:spec_presen}.
    }
\label{fig:spc_line_nonmer}
\end{figure}

\begin{figure*}
    \centering
    \includegraphics[width=\textwidth,clip]{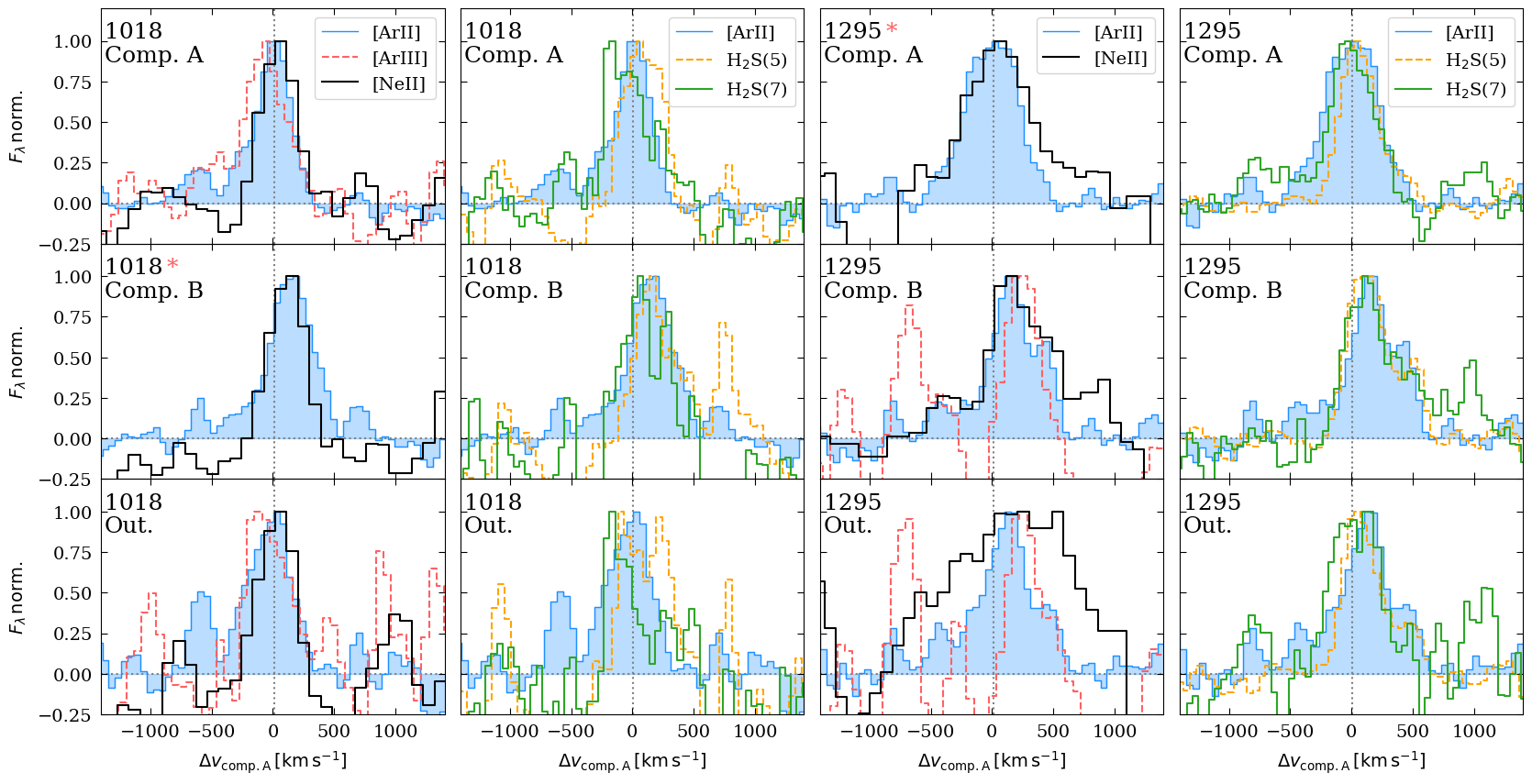}
    \caption{
    Similar to Fig. \ref{fig:spc_line_nonmer} but for the two mergers COSMOS~1018 (with AGN) and COSMOS~1295. In these cases, we also show the warm molecular hydrogen rotational line transitions H$_{2}$S(5) and H$_{2}$S(7), respectively.  We use a red star to indicate the non-detection of [\ion{Ar}{iii}].
    }
\label{fig:spc_line_mer}
\end{figure*}

\begin{figure}
    \centering
    \includegraphics[width=\columnwidth,clip]{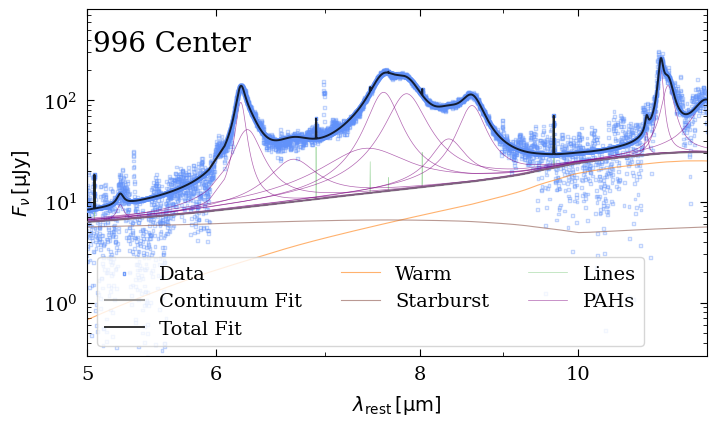}
    \caption{
    Example \texttt{CAFE} fit for COSMOS~996 (center spectrum). Shown are several fit components including continuum (stellar emission and warm dust continuum), emission lines, and PAH bands. Note that the PAH$3.3\,{\rm \mu m}$ band as well as the narrower fine-structure lines and H$_2$ transitions are fit separately.    
    }
\label{fig:cafe_examp}
\end{figure}

\section{Measurement of Mid-IR Emission lines and Bands}\label{sec:analysis}

\subsection{Definition of Apertures}\label{subsec:aper_ana}
In this paper, we provide a comparison of the mid-IR spectral properties in two distinct regions of the galaxies: their central core (here referred to as ``center'') and their diffuse outskirts (here referred to as ``outskirt''). In the case of the two mergers, we extract two central spectra for each component and one spectrum for the extended emission surrounding the components (``outskirt''). Visualizations of the apertures are shown in Fig. \ref{fig:nircam_rgb}. While we here focus on these distinct regions only, we will leave a detailed analysis of the per-pixel variations of PAHs and other mid-IR features to a forthcoming publication.

The outer elliptical apertures are chosen to optimize the area coverage and the signal-to-noise ratio (S/N). To this end, for each galaxy, we compare the spectra extracted from a series of apertures with increasing radii, thereby fixing the semi-major to minor axis ratio ($a/b$) and position angles (PA).
We find an average outer aperture semi-major of $a=1.4\arcsec$.
For COSMOS~1148, we use $a=1.6\arcsec$ to include the bright node in the southeast.
The apertures for the central region are defined as $a=0.5\arcsec$ and $a/b$ and PA are fixed to the same values as the outer apertures. This value is chosen to match the PSF size at $22\rm\,\mu m$ (Sect. \ref{subsec:data_process}).
For mergers, we place two apertures to include each of the merging components (Fig. \ref{fig:nircam_rgb}; blue and red ellipses).
The outskirt spectra are then defined as the spectra extracted from the larger region excluding the central part.
Fig. \ref{fig:spc_present} shows a gallery of the extracted spectra for these different components. The total integrated spectra (center$+$outskirt) are presented in Appendix~\ref{app:spec_presen}.
The following measurement are then performed in each of these spatial regions. Under this treatment, the central and outskirt regions focus on physical scales of $r\lesssim$3.5~kpc and 3.5~kpc$\lesssim r \lesssim$9.4~kpc, respectively.

\subsection{Fitting of Narrow Mid-IR Spectral Features}\label{subsec:z_sys_linefit}

In the following we describe the fitting procedure for ionized and rotationally excited narrow-line features. The systematic and photometric uncertainties in each of the measurements are quantified by a Monte Carlo (MC) procedure. We generate 100 cube realizations for each galaxy by sampling pixels assuming a Gaussian distribution with a $\sigma$ based on the MRS variance cube. We report the median of the 100 fits as well as the 16th to 84th percentiles for each line of each galaxy.

Several mid-IR emission lines are covered by the MIRI MRS observations (Fig. \ref{fig:spc_present}). These spectral features include Br$\alpha$, [\ion{Fe}{II}]5.340\,$\mu$m (hereafter [\ion{Fe}{II}]), [\ion{Ar}{ii}]6.985\,$\mu$m ([\ion{Ar}{ii}]), [\ion{Ne}{VI}]7.652\,$\mu$m ([\ion{Ne}{VI}]), [\ion{Ar}{III}]8.991\,$\mu$m ([\ion{Ar}{iii}]), and [\ion{Ne}{II}]12.814\,$\mu$m ([\ion{Ne}{ii}]).
Zoomed-in example spectra of some of these lines for two non-mergers are shown in Fig. \ref{fig:spc_line_nonmer} (see Appendix~\ref{app:spec_presen} for other non-mergers). Four galaxies (COSMOS~996, 844, 1757, and 1346, marked with ``D'' in Fig. \ref{fig:nircam_rgb}) show double-peaked [\ion{Ar}{ii}] line profiles indicative of disk rotation.
We also observe five molecular hydrogen pure rotational transition lines, H$_{2}$ 0-0 S(2), S(3), S(4), S(5), and S(7) (H$_{2}$S($j$) hereafter). Fig. \ref{fig:spc_line_mer} shows the zoomed-in spectra of the ionized lines as well as the H$_{2}$S($5$) and H$_{2}$S($7$) molecular hydrogen lines for the two merging galaxies. Some of these lines ([\ion{Ar}{ii}] and H$_{2}$) show an additional emission component which could be induced by the merging process.

The line fluxes are measured by fitting a single (or double in the case of double peaks) Gaussian to the emission lines. The local continuum is modeled and subtracted based on a first order polynomial.
We derive $3\sigma$ upper limits for non-detections based on the RMS noise at the expected wavelength while assuming a line width of 400~$\,\rm km\,s^{-1}$ (median line FWHM from the detections).
Table~\ref{tab:line_fit} in Appendix~\ref{app:line_fit} lists the total fluxes of these mid-IR narrow lines (summed over multiple line components) of each galaxy. 

We also re-evaluated the systemic redshifts of our sources based on the [\ion{Ar}{ii}] lines, which are the brightest (S/N $\gtrsim50$) for all sources\footnote{For galaxies with two kinematic components we use the mean of their line centers as systemic redshift.}. 
Our mid-IR redshift measurements (listed in Table~\ref{tab:sample}) are in good agreement with optical measurements from the literature ($|\Delta z|<0.002$).

\subsection{PAH Decomposition and Fitting}\label{subsec:PAH_fit}

Contrary to the narrow ionized and rotational molecular hydrogen lines, PAH emission is represented by wide bands and can be severely blended by other PAH emission or absorption structures.
For the decomposition and flux measurements of PAHs, we therefore used a different fitting method, specifically using the {\it Continuum And Feature Extraction} \citep[\texttt{CAFE},][]{Marshall_2007,CAFE_2025} code.

We first focus on bright PAH features by fitting the wavelength range $\rm 3\,\mu m<\lambda_{rest}<12~\mu m$. This wavelength range is chosen broad for a better coverage of continuum and to avoid the low sensitivity in MIRI channel 4.
\texttt{CAFE} jointly fits the emission lines together with a continuum comprising a stellar component and thermal dust emission with various temperatures \citep[][]{Lai_2022}. 
Note that the much narrower mid-IR emission lines (Sect. \ref{subsec:z_sys_linefit}) are fit jointly with the PAH emission in \texttt{CAFE}, hence they do not impact the flux measurements of the latter\footnote{However, we found that their fluxes themselves are not reliable. This is probably due to that some of the lines cannot be fitted with a single Gaussian (Sect. \ref{subsec:z_sys_linefit}). }. An example \texttt{CAFE} fit over this wavelength range is shown in Fig. \ref{fig:cafe_examp}.

The PAH3.3\,$\mu$m band is of significantly lower S/N although more spectrally isolated. To avoid degeneracies with the continuum level, we ran \texttt{CAFE} locally in the $\rm 3\,\mu m<\lambda_{rest}<5\,\mu m$ range (Appendix~\ref{app:line_fit}, Fig. \ref{fig:pah33_fit}).
Due to the low S/N, there is no indication of $\sim3\,{\rm \mu m}$ H$_{2}$O ice absorption in our sample, thus this component is not included in the \texttt{CAFE} fitting. Four of our targets have S/N$_{\rm PAH3.3}<3$, for which we obtain $3\sigma$ upper limits by computing the RMS noise after continuum subtraction (based on a first-order polynomial fit through $3\pm0.05\,\rm\mu m$ and $3.5\pm0.05\,\rm\mu m$) and assuming a line width of $\rm 0.039\, \mu m$ \citep[corresponding to the FWHM of the $3.3\,{\rm \mu m}$ line;][]{Lai_2023}.
The fitted PAH fluxes are reported in Table \ref{tab:pah_fit}.

Dust absorption features in the mid-IR are known to be degenerate with other fitting parameters \citep[see discussion in][]{Lai_2024}. Specifically, the dust attenuation caused by $9.7\,\rm\mu m$ silicate absorption, $\tau_{\lambda}\propto\tau_{\rm Si,9.7\rm\mu m}$ 
\citep[hereafter $\tau_{\rm Si}$;][]{Chiar_2006,Spoon_2007,Lai_2020} is largely degenerate with the underlying continuum.
While \texttt{CAFE} can in principle fit these components simultaneously, we found that the low continuum S/N in the wavelength range around $\lambda_{\rm rest}\approx10\,\mu$m ($\lambda_{\rm obs}\sim19.5\,\mu$m) leaves the optical depth (relative to $\tau_{\rm Si}$) poorly constrained during fitting.
We therefore set $\tau_{\rm Si,9.7\rm\mu m}=0$ for the fitting process but {\em post facto} estimate the impact of dust attenuation on the measured quantities as discussed later in Sect. \ref{subsec:dis_dust_att}.

\begin{table*}
\caption{Observed PAH fluxes in $10^{-19}\,{\rm W\,m^{-2}}$.}\label{tab:pah_fit}
\centering
\begin{tabular}{lccccccc}
\hline
\# & PAH3.3 & PAH6.2 & PAH7.7 & PAH8.3 & PAH8.6  & PAH11.3 & $L_{\Sigma \rm PAH}/L_{\rm IR}^{\dagger}$ \\
\hline
\textbf{1018} &  &  &  &  &  &  & 0.02 (0.03)\\
\cline{1-1}
A & $<2.85$ & $7.00^{+0.39}_{-0.13}$ & $30.08^{+3.54}_{-1.65}$ & $5.70^{+0.54}_{-0.41}$ & $4.75^{+0.22}_{-0.23}$ & $10.47^{+0.88}_{-1.07}$ &\\
B & $<2.61$ & $4.48^{+0.09}_{-0.11}$ & $17.41^{+1.01}_{-1.16}$ & $2.97^{+0.24}_{-0.20}$ & $2.67^{+0.11}_{-0.09}$ & $4.07^{+0.36}_{-0.34}$ &\\
Outskirt & $<9.41$ & $12.98^{+0.35}_{-0.24}$ & $49.64^{+2.59}_{-2.96}$ & $9.00^{+0.53}_{-0.71}$ & $5.94^{+0.26}_{-0.38}$ & $9.35^{+1.73}_{-1.38}$ &\\

\textbf{1295} &  &  &  &  &  &  & 0.03 (0.04)\\
\cline{1-1}
A & $<2.40$ & $5.16^{+0.12}_{-0.05}$ & $15.42^{+0.08}_{-0.11}$ & $2.79^{+0.04}_{-0.07}$ & $3.11^{+0.04}_{-0.02}$ & $5.17^{+0.16}_{-0.12}$ &\\
B & $<1.76$ & $3.97^{+0.27}_{-0.09}$ & $11.36^{+0.75}_{-0.38}$ & $1.66^{+0.19}_{-0.24}$ & $2.25^{+0.10}_{-0.12}$ & $2.62^{+0.19}_{-0.25}$ &\\
Outskirt & $<6.19$ & $9.90^{+0.21}_{-0.51}$ & $28.85^{+0.55}_{-0.22}$ & $6.17^{+0.26}_{-0.18}$ & $5.88^{+0.19}_{-0.12}$ & $8.62^{+0.45}_{-0.30}$ &\\

\textbf{1148} &  &  &  &  &  & & 0.06 (0.07) \\
\cline{1-1}
Center & $<1.98$ & $7.37^{+0.14}_{-0.10}$ & $19.98^{+2.15}_{-2.94}$ & $1.29^{+0.15}_{-0.41}$ & $4.00^{+0.07}_{-0.17}$ & $5.62^{+0.22}_{-0.22}$ &\\
Outskirt & $<10.91$ & $25.55^{+0.20}_{-0.26}$ & $65.56^{+0.54}_{-0.33}$ & $6.45^{+0.18}_{-0.22}$ & $16.91^{+0.13}_{-0.14}$ & $13.92^{+0.80}_{-1.09}$ &\\

\textbf{1025} &  &  &  &  &  &  & 0.08 (0.10)\\
\cline{1-1}
Center & $1.55^{+0.14}_{-0.13}$ & $11.15^{+0.03}_{-0.06}$ & $36.95^{+0.23}_{-0.26}$ & $3.65^{+0.03}_{-0.04}$ & $7.34^{+0.02}_{-0.03}$ & $6.53^{+0.09}_{-0.08}$ &\\
Outskirt & $5.27^{+0.30}_{-0.31}$ & $33.78^{+0.06}_{-0.06}$ & $105.22^{+0.16}_{-3.39}$ & $10.96^{+0.03}_{-0.52}$ & $21.97^{+0.03}_{-0.17}$ & $14.47^{+0.26}_{-0.23}$ &\\

\textbf{996} &  &  &  &  &  &  & 0.08 (0.10)\\
\cline{1-1}
Center & $2.08^{+0.28}_{-0.43}$ & $22.99^{+0.11}_{-0.76}$ & $79.25^{+0.44}_{-13.74}$ & $7.55^{+0.07}_{-0.62}$ & $15.62^{+0.05}_{-0.12}$ & $14.31^{+0.21}_{-0.29}$ &\\
Outskirt & $7.11^{+1.20}_{-1.41}$ & $64.15^{+1.85}_{-1.25}$ & $183.80^{+26.24}_{-11.15}$ & $17.61^{+3.63}_{-1.29}$ & $42.94^{+1.57}_{-0.50}$ & $37.52^{+1.01}_{-0.80}$  &\\

\textbf{844} &  &  &  &  &  &  & 0.03 (0.04)\\
\cline{1-1}
Center & $1.33^{+0.19}_{-0.20}$ & $8.16^{+0.16}_{-0.26}$ & $28.92^{+1.25}_{-5.00}$ & $<3.58$ & $5.74^{+0.07}_{-0.20}$ & $5.81^{+0.11}_{-0.20}$ &\\
Outskirt & $2.29^{+0.69}_{-0.44}$ & $22.01^{+0.26}_{-0.14}$ & $61.85^{+0.66}_{-0.18}$ & $<7.08$ & $13.81^{+0.26}_{-0.07}$ & $9.34^{+0.43}_{-0.38}$ &\\

\textbf{1757} &  &  &  &  &  &  & 0.04 (0.05)\\
\cline{1-1}
Center & $1.27^{+0.13}_{-0.12}$ & $7.63^{+0.17}_{-0.06}$ & $27.63^{+1.79}_{-3.38}$ & $3.63^{+0.10}_{-0.41}$ & $5.69^{+0.07}_{-0.06}$ & $5.32^{+0.46}_{-0.34}$ &\\
Outskirt & $4.73^{+0.48}_{-0.46}$ & $23.53^{+0.36}_{-0.10}$ & $74.99^{+0.55}_{-0.25}$ & $8.13^{+0.18}_{-0.19}$ & $16.63^{+0.15}_{-0.07}$ & $14.17^{+0.47}_{-0.28}$ &\\

\textbf{1346} &  &  &  &  &  &  & 0.05 (0.06)\\
\cline{1-1}
Center & $<2.57$ & $4.02^{+0.28}_{-0.13}$ & $16.16^{+0.36}_{-0.56}$ & $2.38^{+0.09}_{-0.16}$ & $3.77^{+0.03}_{-0.04}$ & $5.79^{+0.63}_{-0.65}$ &\\
Outskirt & $<11.49$ & $16.35^{+1.36}_{-1.60}$ & $58.93^{+0.83}_{-0.64}$ & $6.86^{+0.17}_{-0.20}$ & $12.74^{+0.06}_{-0.11}$ & $18.24^{+1.90}_{-1.69}$ &\\

\hline
\end{tabular}
\parbox{\textwidth}{
\smallskip
\textbf{Notes:} $^{\dagger}$ Ratio between total PAH luminosity, $L_{\Sigma \rm PAH}$, and total infrared luminosity, $L_{\rm IR}$ (Table \ref{tab:sample}). The values in the parentheses are corrected for the contribution of PAH$12.6\,{\rm \mu m}$ and PAH$17\,{\rm \mu m}$, which are not covered by MIRI (see text).
}
\end{table*}

\section{Results}\label{sec:results}


\subsection{Hidden AGN Traced by Mid-IR Continuum and Lines}\label{subsec:res_AGN}
\begin{figure}
    \centering
    \includegraphics[width=\columnwidth,clip]{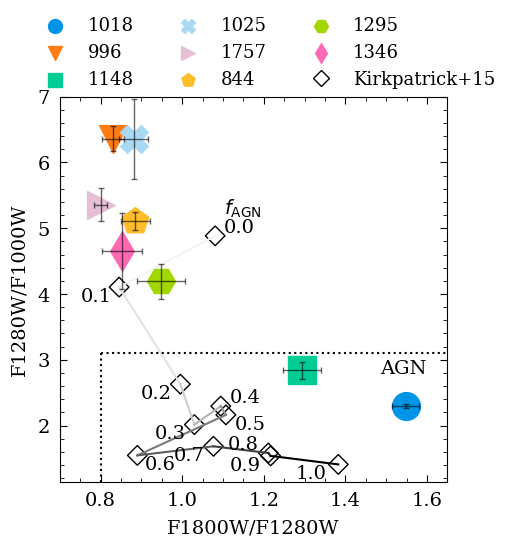}
    \caption{
    Mid-IR color-color diagram based on MIRI photometry in F1000W, F1280W, and F1800W. Our sample is shown in colored large symbols. The open diamond markers show the empirical stacks from \citet{Kirkpatrick_2015} with varying AGN luminosity fractions ($f_{\rm AGN}$, indicated). Dotted black lines empirically separate galaxies with AGN contributions of more than $10\%$ (including two of our targets) from the rest.    
    }
\label{fig:diag_midircolor}
\end{figure}
\begin{figure*}
    \centering
    \includegraphics[width=\textwidth,clip]{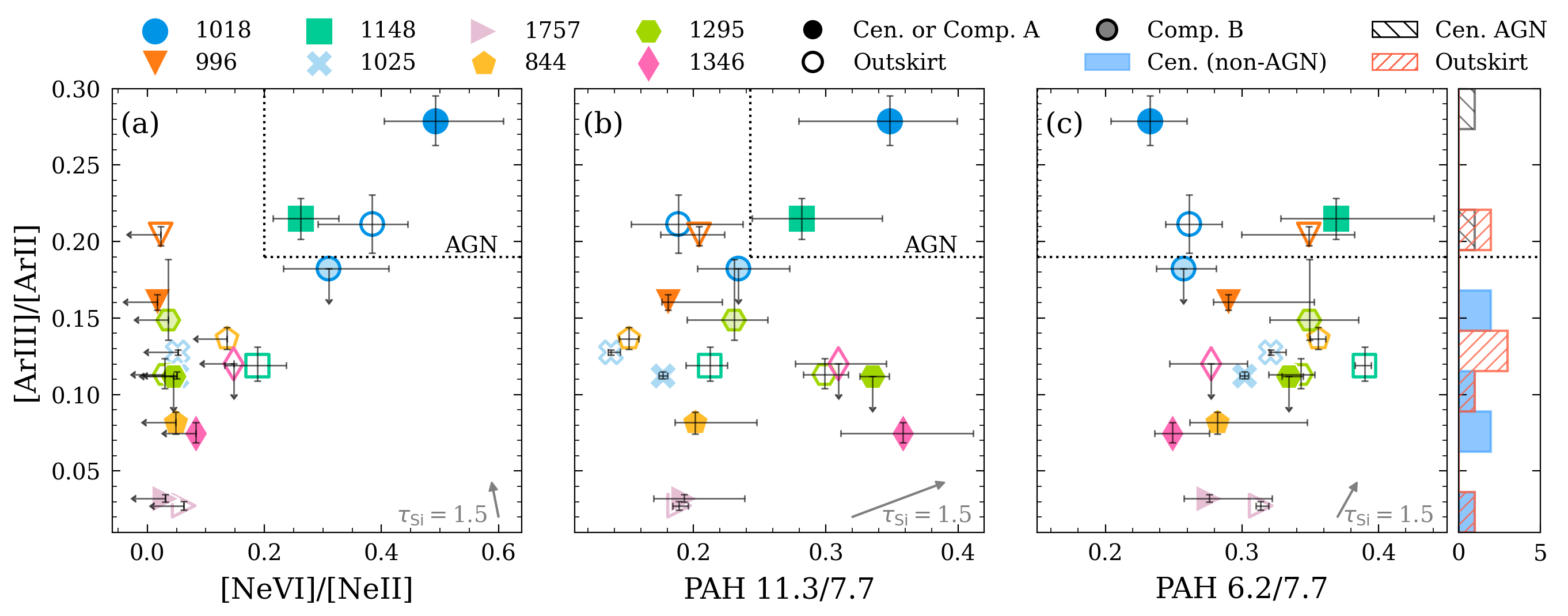}
    \caption{
    [\ion{Ar}{III}]/[\ion{Ar}{II}] (indicator of ionization strength) versus other line ratios including [\ion{Ne}{VI}]/[\ion{Ne}{II}] {\em (a)} PAH 11.3/7.7 {\em (b)}, and PAH 6.2/7.7 {\em (c)}.
    Our galaxies are shown in colored symbols. Measurements for central (outskirt) regions are shown with filled (open) symbols. The secondary components in the merging systems are marked with fainter filling.
    In each panel, the dotted lines mark the empirical separation of AGN from the rest.
    The right-most panel shows a histogram in [\ion{Ar}{iii}]/[\ion{Ar}{ii}] separating central and outskirt regions.
    }
\label{fig:diag_Ar_Ne_pah}
\end{figure*}

Our sample was primarily selected to contain star-forming main-sequence galaxies. COSMOS~1148 is the only target in our sample that shows a nearby ($\sim0.3\arcsec$) X-ray detection in the Chandra COSMOS Legacy Survey catalog \citep[][]{Civano_2016} with $f_{\rm 0.5-10keV}=(1.18\pm0.12)\times10^{-14}\,\rm erg\,s^{-1}\,cm^{-2}$ or $L_{\rm X-ray}=(5.5\pm0.6)\times 10^{43}\,\rm erg\,s^{-1}$. This is a clear sign of AGN activity in that galaxy \citep[][]{Mateos_2012}.

The new MIRI photometric and spectroscopic data allow for an independent view on AGN activity (complementary to UV, optical and X-ray AGN diagnostics) using the mid-IR color and ionized mid-IR diagnostic lines.
Fig. \ref{fig:diag_midircolor} shows an F1800W/F1280W vs. F1280W/F1000W color-color diagram of all galaxies in our sample.
At these redshifts, a higher F1800W/F1280W suggests a rising red continuum and a higher F1280W/F1000W implies stronger PAH6.2$\,\rm \mu$m emission (which is contaminating the F1280W filter).
The diamonds show the position of template AGN spectra from \citet{Kirkpatrick_2015}, which are stacks of galaxies binned by an increasing fraction of AGN contribution to the total luminosity ($f_{\rm AGN}$). The stacked spectra were convolved with the same filters and redshifted to $z=0.93$.
For increasing AGN luminosity we therefore would expect a redder F1800W/F1280W color and a bluer F1280W/F1000W color as PAH dust is dissociated in these environments.
In addition to COSMOS~1148 (confirmed X-ray AGN), we find that also COSMOS~1018 is in the region expected for galaxies with a high AGN luminosity fraction ($f_{\rm AGN}>0.2$) based on these templates.
The morphology of COSMOS 1018A indeed shows an unresolved point-like source indicating central emission from an AGN torus.
Its MIRI/MRS spectrum (component A; Fig. \ref{fig:spc_present}) shows a steepening of the continuum towards longer wavelengths, which is indicative of dust heated by the AGN. Note that a less steep spectrum is found for COSMOS~1148 which could represent different torus dust-to-gas ratio and distribution than COSMOS~1018A \citep[][]{Esparza-Arredondo_2021}. 

In Fig. \ref{fig:diag_Ar_Ne_pah}a, we show the line ratios [\ion{Ne}{VI}]/[\ion{Ne}{II}] vs. [\ion{Ar}{iii}]/[\ion{Ar}{ii}] for our sample. Due to the high ionization potential energy of [\ion{Ne}{VI}] ($126.2\,{\rm eV}$), a detection of this line is a smoking gun for the presence of an AGN. Indeed, both COSMOS~1018 and COSMOS~1148 show strong [\ion{Ne}{VI}] line emission ([\ion{Ne}{VI}]/[\ion{Ne}{II}]$\gtrsim0.2$) and occupy a different phase space in this diagram compared to the other galaxies.
Note that the central regions show a higher ionization compared to the outskirts, which is expected if the AGN affects more the central parts of the galaxies.

\subsection{PAHs in $z\sim1$ Main-Sequence Galaxies}\label{subsec:res_pah}

In the following, we present results involving different PAH band ratios and put our sample in context by comparing it to measurements at lower and higher redshifts.

\subsubsection{PAH Emission in Different Ionizing Environments}

The correlation between PAH ratios and [\ion{Ar}{iii}]/[\ion{Ar}{ii}] are shown in Fig. \ref{fig:diag_Ar_Ne_pah}b and \ref{fig:diag_Ar_Ne_pah}c.
The latter is an indicator of the ionization state of the ISM\footnote{Note that commonly the line flux ratio of [\ion{Ne}{II}] to [\ion{Ne}{III}]$15.555\rm\mu$m is used as ionization state tracer \citep[e.g.,][]{Lai_2022}. However [\ion{Ne}{III}] is redshifted out of the MIRI coverage. We therefore use here the Argon ratio as an alternative tracer of ionization.}.
The AGN (COSMOS~1018 and 1148 at high [\ion{Ar}{iii}]/[\ion{Ar}{ii}]) tend to have larger PAH 11.3/7.7 ratios, similar to COSMOS~1295 (merger) and COSMOS~1346 (which both have lower [\ion{Ar}{iii}]/[\ion{Ar}{ii}]). Note that COSMOS~1346 has the highest PAH 11.3/7.7 ($0.36_{-0.04}^{+0.05}$) in our sample, yet its nature is unclear given the current observations (see Sect. \ref{subsec:dis_agn_merger}).
The PAH 11.3/7.7 is related to the fraction of ionized PAHs, indicating that these galaxies are more dominated by {\em neutral} PAHs (e.g., Sect. \ref{subsub:pah_diag}). The PAH 6.2/7.7 ratio (sensitive to molecule size) does not show a clear trend between AGN and non-AGN nor the ionization state traced by the argon line ratio.

\subsubsection{PAH ratios and warm H$_2$}\label{subsub:pah_h2}
\begin{figure*}
    \centering
    \includegraphics[width=0.8\textwidth,clip]{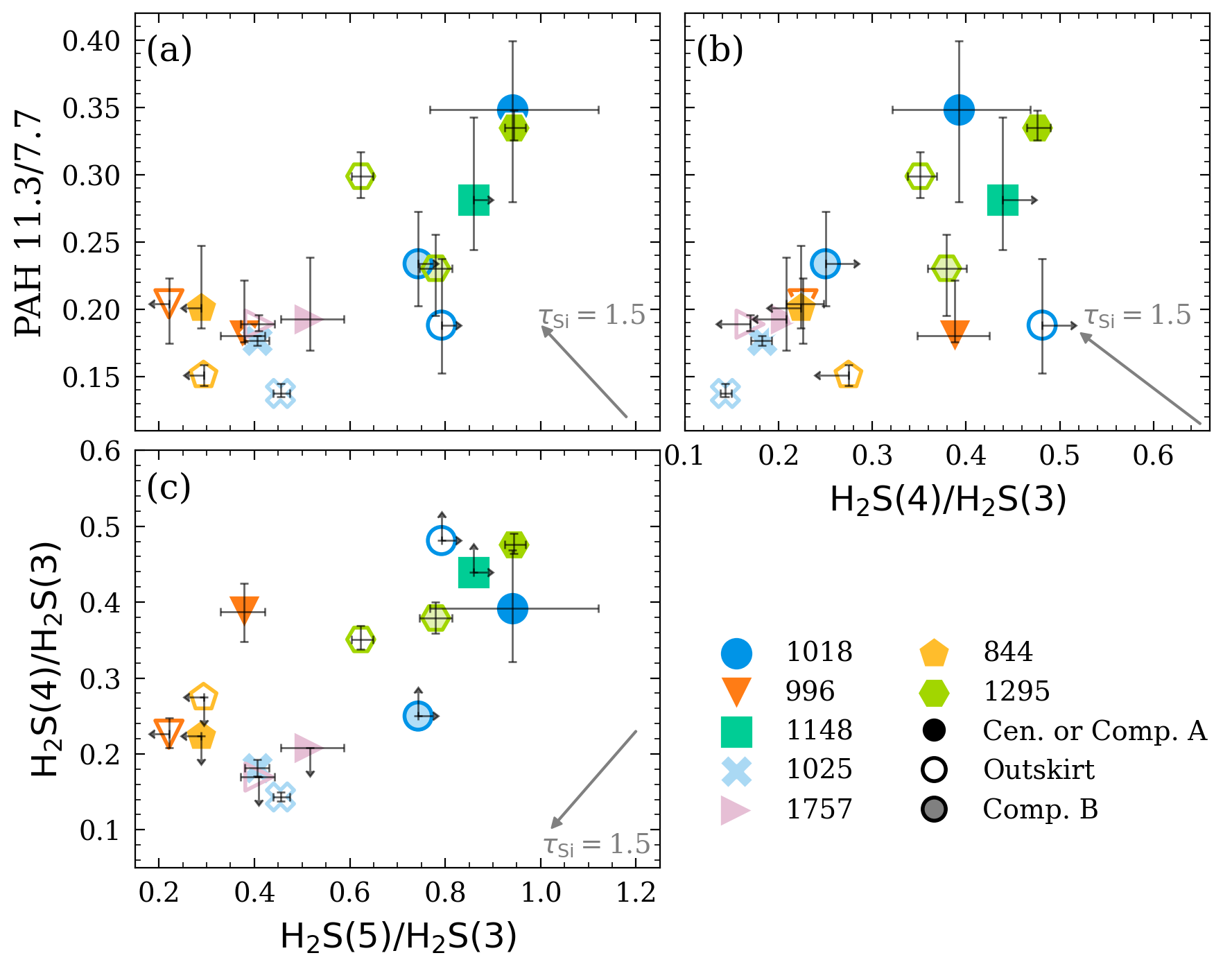}
    \caption{
    Comparison of PAH ratios and H$_2$ warm molecular hydrogen line ratios.
    {\em (a)}: H$_{2}$S(5)/H$_{2}$S(3) versus PAH 11.3/7.7.
    {\em (b)}:  H$_{2}$S(4)/H$_{2}$S(3) versus PAH 11.3/7.7.
    {\em (c)}: H$_{2}$S(5)/H$_{2}$S(3) versus H$_{2}$S(4)/H$_{2}$S(3).
    The symbols and dust arrow are the same as in Fig. \ref{fig:diag_Ar_Ne_pah}. COSMOS 1346 has no detection of H$_2$ lines thus is excluded from this diagnostics. Both AGN and mergers show enhanced H$_2$ line ratios indicative of shocks.
    }
    \label{fig:diag_h2pah}
\end{figure*}

The harsher environment that exists in galaxies hosting AGN and the shocked regions of merging systems has implications on dust and gas. In our MIRI MRS data, we observe five pure rotational H$_{2}$ emissions which imply the existence of warm molecular hydrogen gas ($T\gtrsim100\,$K). In addition to the radiative heating from young stars, non-radiative process such as shocks could lead to the excitation of the H$_{2}$ lines \citep[][]{Cluver_2010,Guillard_2012,Togi_2016,Kristensen_2023,Appleton_2023}. As shown in Fig. \ref{fig:diag_h2pah}, the AGN and mergers in our sample show detections in {\em all} high order transitions (H$_{2}$S($j>4$)). Specifically, we find an increased PAH 11.3/7.7 ratio with increased H$_{2}$S(5) to H$_{2}$S(3) line flux ratio. This implies that not only a harsh radiation environment (traced by [\ion{Ne}{VI}]) can lead to a higher fraction of neutral PAHs but also non-radiative excitation (e.g., by shocks) as traced by high-$j$ H$_2$ rotational transitions. We will further discuss the behavior of PAHs in AGN and mergers in Sect. \ref{subsec:dis_agn_merger}.

\subsubsection{PAH Band Ratios as Tracers of Charge and Size}\label{subsub:pah_diag}

\begin{figure*}
    \centering
    \includegraphics[width=\textwidth,clip]{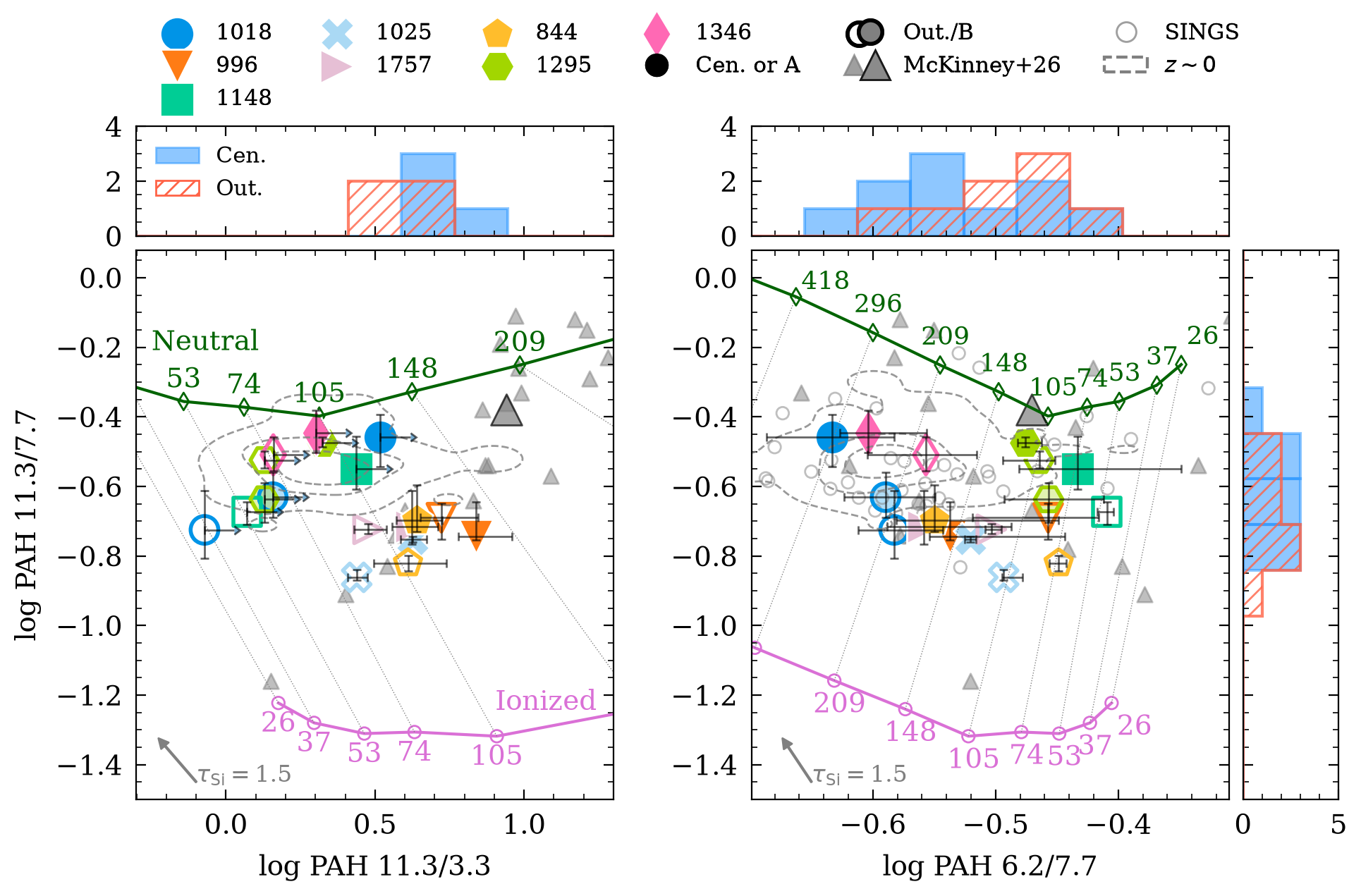}
    \caption{
    PAH ratio diagnostic diagrams for PAH 11.3/3.3 versus PAH 11.3/7.7 ({\em left panel}) and PAH 6.2/7.7 versus PAH 11.3/7.7 ({\em right panel}). The symbols are the same as in Fig. \ref{fig:diag_Ar_Ne_pah}. We overlay models of neutral (green diamonds) and ionized (pink circles) for various numbers of carbon atoms ($N_{\rm C}$) on each panel (see text for derivation). The blue and red histograms show the PAH ratios (excluding limits) of our sample of center and outskirt, respectively.
    We also show measurements from typical $z\approx0$ galaxies from the SINGS sample where available \citep[empty circles;][]{Smith_2007}, a collection of $z\approx0$ galaxies with $L_{\rm IR }>10^{10}\,L_{\odot}$ including GOALS \citep[dashed contours][]{Stierwalt_2013, Stierwalt_2014,Inami_2018, Lai_2020, McKinney_2021a}, as well as $z\approx1.7$ (U)LIRGs \citep[gray triangles;][]{McKinney_2026}.
    }
    \label{fig:diag_pahpah}
\end{figure*}

In Fig. \ref{fig:diag_pahpah} we show the ratio diagnostics of PAH 11.3/3.3 versus PAH 11.3/7.7 and PAH 6.2/7.7 versus PAH 11.3/7.7.
The variation of PAH ratios are related to the PAH grain properties specifically their molecular sizes ($N_{\rm C}$) and charges.  This can be quantified further by models of \citet{Draine_2021}. These model spectra are generated using the \citet{Bruzual_2003} stellar radiation field with an intensity parameter\footnote{The dimensionless intensity parameter $U$ is defined as the ratio between the radiation energy density of the starlight and a reference energy density for standard grains \citep[for details see][]{Draine_2021}. Under this treatment, different incident spectra with the same $U$ will result in the same dust emission. The PAH ratios are not sensitive to $U$ \citep[][]{Draine_2021}. The results with $\log(U)=1$ apply to $\log(U)\lesssim3$.}
$\log(U)=1$ and different average $N_{\rm C}$.
The model PAH fluxes are measured from the model spectra using the same fitting approach as used for the observations.
Fig. \ref{fig:diag_pahpah} includes these models for both neutral (green diamonds) and ionized (pink circles) grains. The molecule sizes (number of carbon atoms, $N_{\rm C}$) are given for each data point; PAH 6.2/7.7 and PAH 11.3/3.3 are related to $N_{\rm C}$ and PAH 11.3/7.7 probes the ionization of the grains.
Note that the normalization of both charge and size are model dependent, hence here we only focus on relative differences of these parameters in our sample.

The PAH3.3\,$\mu$m emission is only detected in half of the galaxies (for both center and outskirt). We find median line flux ratios (for detections) at the galaxy center and outskirt of PAH 11.3/3.3$_{\rm cen}=4.3_{-0.1}^{+1.4}$ and PAH 11.3/3.3$_{\rm out}=3.5_{-0.7}^{+1.2}$, respectively.
For the other line ratios we find:
PAH 11.3/7.7$_{\rm cen}=0.23_{-0.05}^{+0.11}$,
PAH 11.3/7.7$_{\rm out}=0.20_{-0.04}^{+0.09}$,
PAH 6.2/7.7$_{\rm cen}=0.29_{-0.03}^{+0.06}$,
and PAH 6.2/7.7$_{\rm out}=0.33_{-0.05}^{+0.02}$.
The errors indicate 16th to 84th percentiles and both merger components are combined as center regions in these statistics.
The histogram panels of Fig. \ref{fig:diag_pahpah} show the distributions of PAH ratios measured in the center and outskirt regions excluding limits.

While our sample spans a large distribution of grain sizes, it is skewed towards more neutral grains. AGN tend to be located towards models indicating more neutral gains. In Sect.~\ref{subsec:dis_dust_att} we further discuss the potential impact of the dust attenuation which may reduce the separation between AGN and star formation galaxies. We also find indications that the center regions are skewed towards larger PAH 11.3/3.3, larger PAH 11.3/7.7, and smaller PAH 6.2/7.7 ratios compared to the outskirts. These results are limited by small-number statistics and should be interpreted with caution. We will discuss the significance of these results in Sect. \ref{subsec:dis_pah_spatial}.

\subsubsection{Our Sample in Context}\label{subsub:res_compare}

To put our sample into a larger context, we include various literature samples in Fig. \ref{fig:diag_pahpah}.
These include local star-forming galaxies \citep[SINGS, $L_{\rm IR}\lesssim10^{10}\,L_{\odot}$,][]{Kennicutt_2003,Smith_2007}\footnote{Note that $L_{\rm IR}$ for the SINGS sample was derived by integration over $3-1100\,\mu$m, while we use $8-1000\,{\rm \mu m}$. However, this only leads to negligible differences of a few per-cent. Furthermore, we were not able to find PAH$3.3\rm \mu m$ measurements for the SINGS sample, hence we do not include this sample in some figures.}, local luminous infrared galaxies \citep[$L_{\rm IR}\gtrsim10^{10}\,L_{\odot}$, including GOALS (U)LIRGs; referred to as ``$z\sim0$ sample'' throughout this work,][]{Stierwalt_2013,Stierwalt_2014,Inami_2018,Lai_2020,McKinney_2021a}, and high-redshift $0.6<z<2.46$ (U)LIRGs with $\left<z\right>=1.7$ \citep[][]{McKinney_2026}.
Comparing these samples to our galaxies we find
{\it (i)} the PAH 11.3/3.3 ratio of our four detections are in between the values from the $z\sim0$ sample ($2.1_{-0.6}^{+0.9}$) and the $z\approx1.7$ (U)LIRG ($8.7_{-4.9}^{+7.8}$) sample;
{\it (ii)} the PAH 11.3/7.7 ratio of our sample is comparable to the SINGS ($0.28_{-0.06}^{+0.18}$) and the $z\sim0$ sample ($0.29_{-0.03}^{+0.03}$);
{\it (iii)} the PAH 11.3/7.7 ratios of our sample are lower than the $z\approx1.7$ (U)LIRGs ($0.4_{-0.2}^{+0.3}$);
{\it (iv)} the PAH 6.2/7.7 ratios of our sample agree with the ratios of the SINGS ($0.28_{-0.05}^{+0.10}$), the $z\sim0$ ($0.26_{-0.02}^{+0.03}$), and the $z\approx1.7$ (U)LIRG (with a larger scatter, $0.3_{-0.1}^{+0.2}$) samples.
It is important to note that these samples have different total IR luminosities and can include AGN.  The impact of these to contributions will be discussed in Sect. \ref{subsec:dis_pah_evol}.

\begin{figure}
    \centering
    \includegraphics[width=\columnwidth,clip]{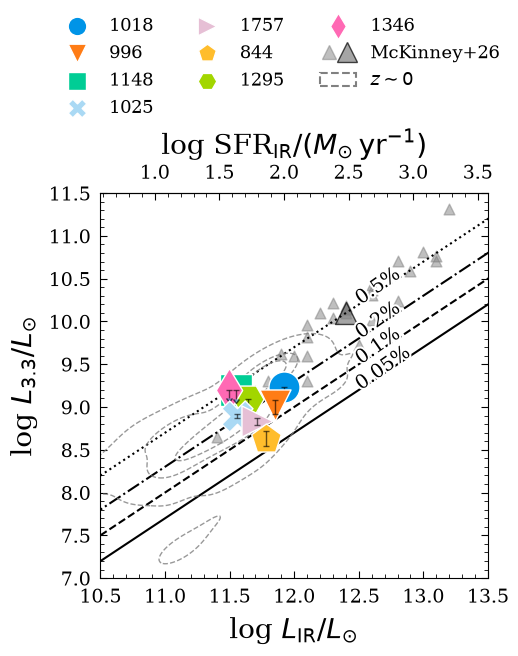}
    \caption{PAH3.3\,$\mu$m luminosity ($L_{\rm 3.3}$) versus $L_{\rm IR}$ and SFR$_{\rm IR}$ \citep[][]{Inami_2022}. The $L_{\rm 3.3}$ of our sample includes both center and outskirt. We also compare with local (dashed contours) and higher redshift (U)LIRG (triangles). We mark $L_{\rm 3.3}/L_{\rm IR}=0.5\%$ $0.2\%$, $0.1\%$, and $0.05\%$ in dotted, dash-dotted, dashed, and solid black lines, respectively. 
    }
\label{fig:diag_lir_pah33}
\end{figure}

\subsubsection{Contribution of PAHs to Total IR Luminosity}
The PAH3.3\,$\rm \mu m$ luminosity has been found to correlate with SFR \citep{Peeters_2004,Shipley_2016,Xie_2019,Lai_2020}. In Fig. \ref{fig:diag_lir_pah33}, we show the $L_{\rm 3.3}$, accounting for both center and outskirt, versus $L_{\rm IR}$, i.e., SFR$_{\rm IR}$ (derived using the \citet{Inami_2022}). We also include local and higher redshift (U)LIRGs for comparison (see Sect. \ref{subsub:res_compare} for details).  For our sample we find $L_{\rm 3.3}/L_{\rm IR}=0.1\%-0.2\%$ (median $0.14\%$), which is consistent with the $L_{\rm 3.3}$-SFR relation of local samples. However, the $z\approx1.7$ McKinney et al. sample is offset to higher ratios, spanning $L_{\rm 3.3}/L_{\rm IR}=0.2\%-0.7\%$ at up to an order of magnitude higher total IR luminosity. This trend may be due to a different stage of evolution (Sect. \ref{subsec:dis_pah_evol}).

Next we compute the fraction of PAH band {\em luminosity} to the total IR luminosity $L_{\Sigma \rm PAH}/L_{\rm IR}$. Note that this is different from $q_{\rm PAH}$, which commonly denotes the {\em mass fraction} of PAHs to total dust mass.
We derived the $L_{\Sigma \rm PAH}$ of each galaxy by summing the PAH emissions from 3.3$\,\rm \mu m$ to 11.3$\,\rm \mu m$. The PAHs at 12.6$\,\rm \mu m$ and 17$\,\rm \mu m$ (not covered by our observations) could contribute a non-negligible fraction of up to $\sim18\%$ to the total PAH luminosity.
We correct for this fraction by using $L_{\Sigma \rm PAH,corr}=L_{\Sigma \rm PAH}+1.13\times L_{6.2}$ according to the empirical median relation derived from local star-forming galaxies \citep{Smith_2007}. The total IR luminosity was derived from the best-fit \texttt{CIGALE} SED \citep{Faisst_2026_UVbump}.
We find $L_{\Sigma \rm PAH}/L_{\rm IR}\approx 2-8\%$ (or $3-10\%$ corrected), which is consistent with the lower end of the distribution ($\sim3-16\%$) found for local star-forming galaxies \citep[][]{Smith_2007}.

\section{Discussion}\label{sec:discuss}

\subsection{Impact of Mid-IR Dust attenuation}\label{subsec:dis_dust_att}

As described in Sect. \ref{subsec:PAH_fit}, we assume $\tau_{\rm Si}=0$ for the \texttt{CAFE} fitting to mitigate degeneracies in the fitting process. Here, we discuss the implications of this choice, namely the impact of possible non-zero mid-IR attenuation, $\tau_{\rm Si}>0$, caused by the 9.7~$\rm \mu m$ silicate absorption. Note that the PAH emission of the comparison samples included in this work are corrected for dust attenuation with a $\tau_{\rm Si}\sim0.4$, $\tau_{\rm Si}\sim1$, and $\tau_{\rm Si}\sim1.5$ for SINGS, $z\sim0$ (U)LIRGs, and $z\approx1.7$ ULIRGs, respectively \citep[][]{Smith_2007,Stierwalt_2014,McKinney_2026}.  

As shown in Appendix~\ref{app:atten_check}, we can set an independent approximate limit on the silicate absorption for our sample of $\tau_{\rm Si}<0.3$ with a possible maximal conservative value of $\tau_{\rm Si}<1.5$.
In summary, we derive these limits using two approaches: first, by comparing our spectra with empirical templates that span a range of silicate absorption strengths, and second, by independently estimating the silicate absorption from the spectra using a polynomial fit.
In addition, \citet{McKinney_2026} measured a $\tau_{\rm Si}\sim1$ for the $z\approx1.7$ (U)LIRG sample at significantly higher $L_{\rm IR}$ and high gas density (indicated by the H$_2$O ice absorption). We would expect a lower $\tau_{\rm Si}$ for our sample, which is another verification of the low limits on $\tau_{\rm Si}$ that are assumed here.

We can now discuss the implication of a non-zero mid-IR attenuation on the various results presented in the previous sections.
\citet{Lai_2024} showed that various PAH ratios can be affected by up to $\sim50\%$ when $\tau_{\rm Si}$ is increasing from 0 to 1.5.
Assuming a mixed geometry of the PAH emissions and dust, the intrinsic emission is attenuated by $(1-e^{-\tau_{\lambda}})/\tau_{\lambda}$. The optical depth $\tau_{\lambda}$ of the total dust attenuation can be derived from $\tau_{\lambda}=\tau_{\rm Si}\cdot C_{\rm 9.7\mu m}(\lambda)$, where $C_{\rm 9.7\mu m}(\lambda)$ is the attenuation curve normalized to its value at $\rm 9.7\mu m$. Here we use the \citet{Ossenkopf_1992} attenuation curve, which is also used by \texttt{CAFE}, to estimate the attenuation at the wavelength of each spectral feature.
In Fig. \ref{fig:diag_pahpah}, we show the impact of attenuation on the various PAH ratios assuming the worst-case scenario of $\tau_{\rm Si}=1.5$ (note that more realistic values are $\tau_{\rm Si}\lesssim0.3$).
The arrow denotes how the PAH ratio would change under the correction of the {\em observed} fluxes.
Specifically, we find correction values defines as
$\Delta \rm PAH_{int-obs} (\lambda_{1}/\lambda_{2})=\log PAH_{\rm intrinsic} (\lambda_{1}/\lambda_{2})-\log PAH_{\rm obs, median} (\lambda_{1}/\lambda_{2})$ of 
$0.13\,{\rm dex}$, $-0.13\,{\rm dex}$, and $-0.02\,{\rm dex}$ for $\Delta \rm PAH_{int-obs}(11.3/7.7)$, $\Delta \rm PAH_{int-obs}(11.3/3.3)$, and $\Delta \rm PAH_{int-obs}(6.2/7.7)$, respectively.
The typical logarithmic measurement uncertainties for the same ratios are $0.10\,{\rm dex}$, $0.15\,{\rm dex}$, and $0.07\,{\rm dex}$, respectively, suggesting a non-negligible impact of dust attenuation on our PAH ratios.
However, assuming the more realistic attenuation case of $\tau_{\rm Si}\lesssim0.3$ suggests only a modest impact on the PAH ratios.

Similarly, the narrow emission line ratios may be affected by dust attenuation. 
Applying a similar calculation as above, we find that attenuation does not affect [\ion{Ne}{VI}]/[\ion{Ne}{ii}] and has minimal impact on [\ion{Ar}{iii}]/[\ion{Ar}{ii}] (see arrow in Fig. \ref{fig:diag_Ar_Ne_pah}). 
As for the H$_{2}$ line ratios, our diagnostic involves H$_{2}$S(3) at rest-frame 9.6649$\,\rm \mu m$ closer to to the peak of absorption (see arrow in Fig. \ref{fig:diag_h2pah}). 
Assuming the worst-case scenario of $\tau_{\rm Si}=1.5$, we find corrections of $\Delta$H$_{2}$S(5)/H$_{2}$S(3) $\approx-0.18\,{\rm dex}$ and $\Delta$H$_{2}$S(4)/H$_{2}$S(3) $\approx-0.13\,{\rm dex}$ which are $\times2-3$ larger than the measurement uncertainties ($0.07\,{\rm dex}$ and $0.04\,{\rm dex}$, respectively).
Again, assuming a more realistic value of $\tau_{\rm Si}\lesssim0.3$ would change the impact to minimal.

\subsection{Spatial PAH Variations}\label{subsec:dis_pah_spatial}

MIRI/MRS provides the necessary spatial resolution to investigate variations of PAH emission across the galaxies close to cosmic noon for the first time. In this section, we discuss variations in PAH emission and ionization from the center ($r\lesssim$3.5~kpc) to the outskirts (3.5~kpc$\lesssim r\lesssim$9.4~kpc) of the galaxies.

Taking the results in Sect. \ref{subsec:res_pah} at face value (excluding the two AGN and non-detections), we find differences in the PAH ratios, defined as
$\Delta \rm PAH_{cen-out} (\lambda_{1}/\lambda_{2})=\log  PAH_{center} (\lambda_{1}/\lambda_{2})-\log PAH_{outskirt} (\lambda_{1}/\lambda_{2})$,
of $0.05\,{\rm dex}$, $0.13\,{\rm dex}$, and $-0.05\,{\rm dex}$, for $\Delta \rm PAH_{\rm cen-out}(11.3/7.7)$, $\Delta \rm PAH_{\rm cen-out} (11.3/3.3)$, and $\rm \Delta PAH_{\rm cen-out}(6.2/7.7)$, respectively.
Note that these difference will be enhanced if we assume that the centers are {\em more} dust attenuated ({\em i.e.} larger $\tau_{\rm Si}$) than the outskirts. 

To quantify the statistical significance of these differences, we performed a MC-based two-sample Kolmogorov-Smirnov (K-S) test for each PAH ratio difference for each galaxy separately.
To this end, we assume the observed $\Delta \rm PAH_{cen-out} (\lambda_{1}/\lambda_{2})$ to be a Gaussian distribution with a $\sigma$ equal to the observational uncertainties added in quadrature. The assumed null hypothesis of no difference between center and outskirt is represented by a Gaussian distribution with zero mean and the same $\sigma$.
We then randomly sample $n_{\rm draw} = 500$ from these two distribution and compute the two-sample K-S test. This is repeated $100$ times for each PAH ratio.
For most targets, $>80\%$ of the simulations reject the null hypothesis ($p$-value $<0.05$) suggesting overall a statistically significant difference in the PAH ratios between spatial centers and outskirts.

In summary, the central regions tend toward larger PAH 11.3/3.3 and 11.3/7.7 ratios but smaller 6.2/7.7 ratios compared to the outskirts. PAH 6.2/7.7 and PAH 11.3/3.3 probe the average size of the PAH molecules, and PAH 11.3/7.7 is related to the charge of PAH molecules.
Overall, this indicates a reduction in the average $N_{\rm C}$ ({\em i.e.} smaller molecules) and a decrease in the neutral fraction of dust molecules from the central region to the outer diffuse ISM.
Note that this trend from inside out is comparable to the trends of the AGN in our sample, which are described by more neutral grains. However, we caution that the results are based on a small sample of eight galaxies. Larger samples are required to verify these results.

{\em How does the surrounding radiation field impact the properties of dust molecules?}
If we expect the radiation field to become weaker at the outskirts of galaxies, this would lead to radial changes in the PAH ratios, as the radiation field plays an important role in shaping PAH emission \citep{Baron_2025}.
For example, \citet{Lai_2022} reported an anti-correlation between [\ion{Ne}{III}]/[\ion{Ne}{II}] and distance from the galaxy center in a nearby LIRG. Excluding the nuclear AGN, the authors found a correspondingly higher PAH 6.2/7.7 ratio at larger radii. This suggests the destruction, dissociation, or removal of smaller PAH molecule in harder radiation fields towards the center.
As shown in Fig. \ref{fig:diag_Ar_Ne_pah}, we find no evidence for a clear difference in the [\ion{Ar}{iii}]/[\ion{Ar}{ii}] line ratio between the centers and outskirts of our galaxies.
There could be several reasons for this. First the argon ratio may not trace very strong radiation fields (such as [\ion{Ne}{III}]/[\ion{Ne}{II}])\footnote{Ionization potential of [\ion{Ne}{iii}] and [\ion{Ar}{iii}] is 41\,eV and 27.6\,eV, respectively}. Second, the galaxy studied by \citet{Lai_2022} hosts a strong AGN, enhancing the radiation field within the small physical region of their study, $\lesssim1\,$kpc. However, the central aperture studied in our case is larger ($r\lesssim3.5\,$kpc), therefore may include significant contribution from the host galaxy which could dilute a potential impact of the central AGN. Third, our galaxies may exhibit significant star formation in their disks, which would flatten the radial dependence of the radiation field (for example, COSMOS~884 and 996 show higher [\ion{Ar}{iii}]/[\ion{Ar}{ii}] ratios in the outskirts). 
We note that the two AGN in our sample do show {\em higher} [\ion{Ne}{iv}]/[\ion{Ne}{II}] line ratios in the centers, relatively consistent with the study by \citet{Lai_2022}.

Lastly, we mention that PAHs properties are found to be correlated with metallicity. Specifically, PAH sizes are smaller in low metallicity environments \citep[e.g.][]{Smith_2007,Lai_2025}, which may be attributed to different dust formation pathways at different metallicities \citep[][]{Shivaei_2017,Whitcomb_2024,Shivaei_2024}.
Generally, we would expect negative metallicity gradients in these prototypical massive main-sequence galaxies \citep[e.g.,][]{Gillman_2022}. However, we believe that the metal abundances would still be too high to impact the PAH size distribution significantly in these evolved massive galaxies.
Future IFU optical spectroscopy would be needed to test this picture further.

\subsection{Impact of Harsh and Dynamic Environment: AGN and Mergers}\label{subsec:dis_agn_merger}

Both AGN and mergers are known to have diverging PAH ratios compared to normal star-forming galaxies. They typically show lower PAH 6.2/7.7 (larger $N_{\rm C}$) and higher PAH 11.3/7.7 (more neutral or less charged) ratios \citep[e.g.,][]{Jochims_1994,Singh_2025,Maragkoudakis_2020}. Some studies attribute the latter to the ``deionization'' of charged PAHs in high radiation environments \citep[e.g.,][]{Smith_2007,ODowd_2009,Diamond-Stanic_2010,Zhang_Lulu_2022}.
However, the reality is more complicated; first, contamination by host dust torus continuum at $5-8~\rm \mu m$ can lead to an underestimation of PAH $7.7~\rm \mu m$ emission. Secondly, the process itself between the dissociation of charged PAHs and the ionization of neutral PAHs is dynamic, i.e., higher PAH 11.3/7.7 ratios could suggest faster dissociation than ionization in the presence of AGN.

{\em What do we find at $z\approx1$?}
Similar to literature studies, we find that the PAH 11.3/7.7 ratio tends to be higher for both the AGN and mergers in our sample (Sect. \ref{subsec:res_AGN} and Fig. \ref{fig:diag_Ar_Ne_pah}).
This suggests that the bulk of PAHs are less charged in AGN and merging systems. However, we do not find a significant difference in the PAH 6.2/7.7 ratio, i.e., PAH sizes. One interesting finding is that although AGN and mergers in our sample tend to have similarly high PAH 11.3/7.7 ratios compared to the other star-forming galaxies, the [\ion{Ar}{iii}]/[\ion{Ar}{ii}] and [\ion{Ne}{VI}]/[\ion{Ne}{II}] ratio diagnostics hint at different ionization states with harder radiation field only seen in AGN (Fig. \ref{fig:diag_Ar_Ne_pah}).
This further suggests that the elevated PAH 11.3/7.7 ratio in mergers (not showing strong [\ion{Ne}{VI}]) could be shock-induced instead. As shown in Sect. \ref{subsub:pah_h2} (Fig. \ref{fig:diag_h2pah}), the AGN and mergers in our sample have higher H$_{2}$S(5)/H$_{2}$S(3). We stress again that it is non-trivial to simply link the high PAH 11.3/7.7 only to the faster destroy of charged PAHs as the reality should be a dynamical process.


While the PAH 6.2/7.7 ratio does not show a significant difference between AGN, mergers, and other star-forming galaxies in our sample (e.g., Fig. \ref{fig:diag_Ar_Ne_pah}c), the PAH3.3$\,\mu$m band is another indicator of the dust grain size distribution. Tracing some of the smallest PAH molecules, the $3.3\,{\rm \mu m}$ feature can be suppressed significantly in harsh radiation environments such as AGN \citep{Maragkoudakis_2020,Lai_2023}.
In our sample, only four galaxies (which are all non-AGN and non-mergers) are detected in PAH3.3$\,\mu$m, thus confirming that picture (Fig. \ref{fig:diag_pahpah}). The non-detection is indicative of a lower fraction of small dust grains in AGN and merger systems due to the harsher radiation and more dynamic environment. 
In addition, larger dust grains are found to be formed first shortly after a starburst in the case of supernovae dust production \citep[e.g.,][]{Asano_2013}. A starburst related to the merging system could therefore also shift the dust grain size distribution towards larger grain sizes.

Lastly, it is worth to mention that COSMOS~1346 is, next to the AGN and mergers in our sample, the only other galaxy that shows a high PAH 11.3/7.7 ratio. COSMOS~1346 does not show any obvious evidence of AGN nor merger and none of the H$_2$ emissions are detected. Moreover, it is interesting to note that this galaxy has the {\em lowest} SFR and {\em highest} stellar mass measured in the sample, thus could be in the very early process of quenching. Such a scenario is also in agreement with the non-detection of the PAH3.3$\,{\rm \mu m}$ emission, which is shown to depend on the SFRs. However, without further detailed measurements of its star formation history with the aid of optical spectroscopy, its specific PAH properties remain elusive.  

Finally, we discuss the impact of dust attenuation on the results above. As discussed in Sect. \ref{subsec:dis_dust_att}, silicate dust attenuation has an impact on the PAH and H$_2$ line ratios (as indicated by the gray arrows in the figures, representing the maximal conservative $\tau_{\rm Si} = 1.5$). 
If AGN and mergers experience higher dust attenuation (e.g., due to higher gas densities feeding the central black hole or gas compression during the merging event), their H$_2$S(5)/H$_2$S(3) ratios would be more similar to the ones of the other star forming galaxies. On the other hand, the difference in PAH 11.3/7.7 would be more pronounced, strengthening our conclusions of a larger abundance of neutral dust grains in these systems.

\begin{figure*}
    \centering
    \includegraphics[width=0.9\textwidth,clip]{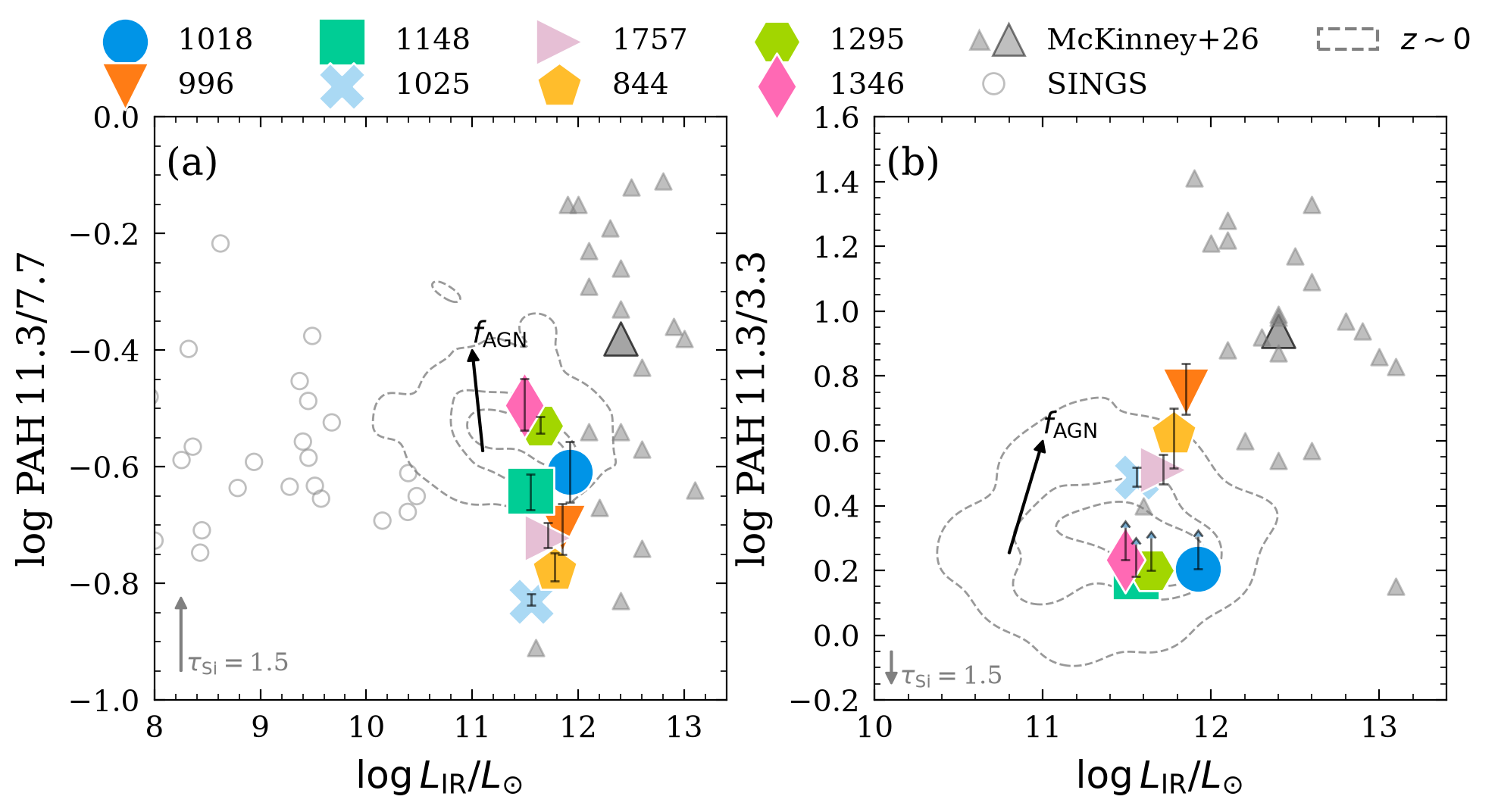}
    \caption{
    Relations of total IR luminosity ($L_{\rm IR}$) with PAH~11.3/7.7 {\em (a)}, and PAH~11.3/3.3 {\em (b)}.
    The symbols are the same as in Fig. \ref{fig:diag_pahpah} (however, we here only show integrated measurement for our sample). The lower limits of the PAH~11.3/3.3 are derived based on the 3$\sigma$ detection limits of PAH3.3\,$\rm \mu m$ non-detections.
    The black arrows indicate the direction of increasing $f_{\rm AGN}$ for the $z\sim0$ sample.
    At a fixed $L_{\rm IR}$, our $z\approx1$ sample is skewed to lower PAH~11.3/7.7 and higher PAH~11.3/3.3 line ratios compared to the $z\approx0$ samples.
    }
    \label{fig:diag_LITpahratio}
\end{figure*}

\subsection{PAH Evolution: Comparison to Literature}\label{subsec:dis_pah_evol}

We would expect the PAH abundance and ratios to change across cosmic time. Galaxies become metal enriched over time \citep{Steidel_2014,Sanders_2021,Papovich_2022,Faisst_2026_metal,WangWuji_2026} while dust mass shows an early buildup and a later decrease with a peak around cosmic noon \citep[e.g.,][]{Faisst_2020,Pozzi_2020,Ferrara_2021}.
Simply speaking, dust is formed early on via supernovae production, then grown in the ISM, and finally produced by AGB stars on longer time scales \citep[e.g.,][]{DellAgli_2017}. The latter will be the main source of dust production  at higher metallicity and later evolutionary stage \citep[e.g., main-sequence galaxies at cosmic noon such as our sample,][]{Valiante_2017}.
The PAH grains size distribution is thought to change based on different production mechanisms, which directly results in different PAH ratios. Starting off as larger grains, smaller PAHs may result from shattering \citep[][]{Jones_1996,Ferrarotti_2006,Asano_2013}.
Previous works suggest that PAHs in ULIRGs at $z\approx2$ exhibit diverse properties, which may be due to the burstiness of their star formation and different cooling mechanisms compared to local populations \citep[e.g.,][]{Lutz_2005,Yan_2007,McKinney_2020}.

The unique sample studied here bridges between the $z\approx2$ (U)LIRGs and local galaxies.
So far, we have compared the PAH ratios alone between different samples at $z=0$ and (U)LIRGs at $z\approx1.7$ (Fig. \ref{fig:diag_pahpah}).
We found that our sample has similar PAH 11.3/7.7 ratios as local galaxies, however, lower ratios compared to $z\approx1.7$ (U)LIRGs. On the other hand, the PAH 6.2/7.7 ratio is similar between all samples studied here.
The PAH 11.3/3.3 ratio is lower in our sample compared to higher-$z$ (U)LIRGs and the contribution of PAHs to the total IR luminosity is at the lower end of the scale ($3-10\%$) compared to other local galaxies ($3-16\%$, the SINGS sample has a median of $10\%$).
In this section, we add another dimension to this discussion, namely the different $f_{\rm AGN}$ and $L_{\rm IR}$ of these galaxy samples.

Fig. \ref{fig:diag_LITpahratio} show the comparison of the PAH 11.3/7.7 and PAH 11.3/3.3 ratios as a function of total IR luminosity. For consistency, we only use the galaxy-integrated values for our sample. First, we see that the samples used here for comparison span more than five orders of magnitude in $L_{\rm IR}$. Our sample lies at the bright-end of the IR luminosity of the local sample but is fainter than the (U)LIRGs at $z\approx1.7$.
To first order, the PAH 11.3/7.7 ratio does not depend on the IR luminosity; however, we find that the $z\approx1.7$ (U)LIRGs sample shows an increased scatter towards higher ratios.
For PAH 11.3/3.3 there is a trend in the sense of an increased ratio at higher IR luminosity. This may be indicative of some true change in the dust grain properties between $z\approx0$ and $z\approx2$.


As discussed in Sect. \ref{subsec:dis_agn_merger}, AGN and mergers can affect the PAH ratios. We expect for our sample a low AGN luminosity fraction ($f_{\rm AGN} < 10\%$, except for the two AGN COSMOS~1018 and 1148), however the other samples compared to here show a larger range (for example, the $z\approx1.7$ (U)LIRGs have $\left<f_{\rm AGN}\right> \sim 19\%$ with a maximum of $63\%$).
Local galaxies with higher AGN luminosity fraction are generally located towards higher PAH ratios as indicated by the arrows in Fig. \ref{fig:diag_LITpahratio}; most of the local galaxies at $\log(\rm PAH 11.3/7.7) > -0.5\,{\rm dex}$ have $f_{\rm AGN} > 30\%$. Similarly, this threshold occurs at $\log(\rm PAH 11.3/3.3) > 0.6\,{\rm dex}$. The trends for the $z\approx1.7$ (U)LIRG sample are more scattered, but generally consistent with an increase of $f_{\rm AGN}$ towards higher ratios.

Controlling for both $f_{\rm AGN}$ and $L_{\rm IR}$, we find that {\em the PAH 11.3/7.7 ratio is indeed lower in our and the $z\approx1.7$ sample compared to the local (U)LIRGs}, while {\em the PAH 11.3/3.3 ratio of our sample is in between the values of the local galaxies and $z\approx1.7$ galaxies.\footnote{We reiterate that the McKinney et al. sample spans a redshift range of $\sim0.6-2.5$, with higher-$z$ sources exhibiting higher $L_{\rm IR}$. }}
Taking this at face value (and noting the caveat of small sample statistics) we can come to conclusion that the PAH 11.3/7.7 ratio increases and the PAH 11.3/3.3 ratio decreases at a fixed $f_{\rm AGN}$ and $L_{\rm IR}$ from $z\sim2$ to $z\sim0$.
This suggests that galaxies towards cosmic noon are more depleted in smaller dust grains, which could be due to the destruction of small grains in dense, highly star-forming environments, and/or a higher production rate of large grains as expected in the early stages following a starburst \citep[see model by][]{Asano_2013}.

A similar mechanism could lead to the differences between our $z\approx1$ sample and the $z\approx1.7$ ULIRG sample. In addition to a higher $f_{\rm AGN}$, these compact dusty statbursts at $z\approx1.7$ are likely forming dust via recent supernovae, yielding primarily larger grains \citep{Asano_2013}, whereas our main-sequence sample is likely dominated by AGB dust formation. As noted by \citet{McKinney_2026}, higher gas densities reduce shattering, preserving larger grains and naturally explaining their elevated PAH 11.3/3.3 ratios relative to our sample.
Distinguishing between a deficit of small grains and an excess of large ones is complex. However, because the $L_{3.3}/L_{\rm IR}$ ratio of  McKinney et al. sample increases rather then declines at higher $L_{\rm IR}$, their PAH $11.3\,{\rm \mu m}$ must be intrinsically more abundant to drive up the  PAH 11.3/3.3 ratio. Indeed, their sample shows $L_{11.3}/L_{\rm IR}\sim3\%$ compared to our $\lesssim1\%$ \citep[which is consistent with local galaxies;][]{Smith_2007,Inami_2018,Lai_2020}. Although the large 0.6~dex uncertainty in $L_{11.3}$ of \citet{McKinney_2026} narrows this gap, the residual offset suggests that large grain production outpaces small grain destruction. 
Furthermore, larger grains have lower UV photoelectric heating efficiencies, potentially increasing the neutral PAH fraction. This elegantly explains the higher PAH 11.3/7.7 ratios in the $z\approx1.7$ ULIRGs compared to our $z\approx1$ sample.

Finally, we mention some caveats.
First, it is important to mention that the local galaxy sample includes a large fraction of galaxies from the GOALS survey \citep{Armus_2009}, which are mostly merging (U)LIRG systems.
The case of COSMOS~1295 (Fig. \ref{fig:diag_Ar_Ne_pah} and \ref{fig:diag_h2pah}) suggests that mergers can have elevated PAH 11.3/7.7 ratios compared to other star forming galaxies (this may be again an effect of recent starbursts as discussed above). Taking this into account would make ours and the local comparison sample more similar.
Second, correcting for a $\tau_{\rm Si} > 0$ would increase the PAH 11.3/7.7 ratio and lessen the difference between our sample and the local comparison sample. The effect on PAH 11.3/3.3 is minimal and the difference between our sample and the $z\approx1.7$ (U)LIRGs is too significant to be affected. Note that the PAHs emission measurements for the galaxy samples at $z\sim0$ and $z\approx1.7$ are corrected for dust attenuation (Sect.~\ref{subsec:dis_dust_att}).

\section{Summary}\label{sec:conclu}

In this paper, we presented the first results from our JWST/MIRI MRS program (GO-4761, PI: A. Faisst) which aims to study the resolved PAH emission of eight galaxies on the star-forming main-sequence at $z\approx1$. The galaxies were spectroscopically selected at $M_{\star}$ of $10^{10.6-11.2}M_{\odot}$ and SFRs of $30-90\,{ M_\odot\,\rm yr^{-1}}$ providing a unique sample to study dust grain properties close to cosmic noon in comparison to local galaxies (Sect. \ref{sec:sample_obs_data}). Two of the galaxies are AGN (Sect. \ref{subsec:res_AGN}) and two are mergers, diversifying the sample. The new JWST/MIRI observations cover all major PAH bands from $3.3-11.3\,{\rm \mu m}$, mid-IR atomic lines (including Br$\alpha$, [\ion{Fe}{II}], [\ion{Ar}{ii}], [\ion{Ne}{VI}], [\ion{Ar}{III}], and [\ion{Ne}{II}]), as well as H$_{2}$ rotational transition lines ($\text{H}_2\text{S}(j), \quad \text{where } j \in \{2, 3, 4, 5, 7\}$). The results of the PAHs and emission lines from the galaxy centers and outskirts are presented in Tables \ref{tab:pah_fit} and \ref{tab:line_fit}. 


The primary observational results of this work are as follows (Sect. \ref{sec:results}):
\begin{itemize}
    \item All major PAH emission features have been detected in our sample, except for the PAH$3.3\,{\rm \mu m}$, which is not detected in the AGN and mergers.

    \item AGN show indications of hard ionizing radiation based on atomic line ratios.  Both AGN and mergers exhibit high-order rotational H$_2$ emission (indicating non-radiative excitation). 

    \item AGN and mergers have elevated PAH 11.3/7.7 ratios (indicating a higher neutral grain fraction). The PAH 6.2/7.7 does not show a clear trend among AGN, merger, and star-forming galaxies.

    \item Spatially, galaxy centers tend to host more neutral and larger grains (higher PAH 11.3/7.7 and lower PAH 6.2/7.7 ratios) than their outskirts. This gradient is most pronounced in AGN.

    \item Compared to local galaxies, our $z\approx1$ sample is skewed towards lower PAH 11.3/7.7 and higher PAH 11.3/3.3 ratios, alongside a lower $L_{\rm \Sigma PAH}/L_{\rm IR}$ fraction ($3-10\%$). Conversely, the comparison $z\approx1.7$ (U)LIRG sample exhibits higher PAH 11.3/7.7 and 11.3/3.3 ratios than our sample.

\end{itemize}



Our results suggest that the harsh environment is destructive to smaller grains (Sect. \ref{subsec:dis_agn_merger}). Comparing across redshifts, we find that accounting for $L_{\rm IR}$ and $f_{\rm AGN}$ is critical, as demonstrated by the correlation between the PAH 11.3/3.3 ratio and $L_{\rm IR}$ (Sect. \ref{subsec:dis_pah_evol}).
The comparison with $z\approx1.7$ (U)LIRGs implies that the dust in our main-sequence $z\approx1$ galaxies is produced by AGB stars and may be processed by shattering, contrary to the rapid large grain production of such compact starbursts (e.g., via supernovae, Sect. \ref{subsec:dis_pah_evol}).
This further suggests a cosmic evolution toward smaller dust grains by $z\sim0$, likely driven by grain shattering. The reduced photoelectric heating efficiency of these larger grains would subsequently increase the neutral PAH fraction.
We note that these results are based on a small sample and larger surveys are necessary to verify them.
 

This paper is the first in a series. Subsequent studies will focus on detailed spatial maps of PAH and fine-structure lines, explore connections to resolved stellar populations, gas kinematics, and investigate the relation between the $2175\,{\rm \AA}$ UV bump and PAH features.

\begin{acknowledgments}
We thank the anonymous referee for their valuable comments and suggestions, which have improved the quality of this manuscript. This work is based on observations made with the NASA/ESA/CSA \textit{James Webb} Space Telescope. The data were obtained from the Mikulski Archive for Space Telescopes at the Space Telescope Science Institute, which is operated by the Association of Universities for Research in Astronomy, Inc., under NASA contract NAS 5-03127 for JWST. These observations are associated with programs JWST-GO-01727 and JWST-GO-04761. Support for US investigators in program JWST-GO-04761 was provided by NASA through a grant from the Space Telescope Science Institute, which is operated by the Association of Universities for Research in Astronomy, Inc., under NASA contract NAS 5-03127.
\end{acknowledgments}

\begin{contribution}
WW, ALF, TL, and KF conducted the data analysis. WW and TL performed the JWST data reduction. WW wrote the majority of the manuscript. ALF, KF, and TL improved the writing. ALF is the PI of the JWST observations used here. All other co-authors contributed to the writing of this manuscript.


\end{contribution}

%
\facilities{JWST}

\software{\texttt{astropy} \citep[][]{astropy_2013,astropy_2018,astropy_2022}; \texttt{photutils} \citep[][]{Bradley_2025a,Bradley_2025b}; \texttt{MPDAF} \citep[][]{Bacon_2016mpdaf,Piqueras_2019mpdaf}; \texttt{q3dfit} \citep[][]{Rupke_2023q3dfit}; \texttt{Jupyter notebook} \citep[][]{kluyver2016jupyter}; \texttt{matplotlib} \citep[][]{Hunter_2007}; \texttt{SciPy} \citep[][]{virtanen2020scipy}; \texttt{NumPy} \citep[][]{harris2020numpy};
\texttt{CAFE} \citep{CAFE_2025}
          }


\appendix

%


\section{Additional Spectra Presentation}\label{app:spec_presen}
\begin{figure*}
    \centering
    \includegraphics[width=\textwidth,clip]{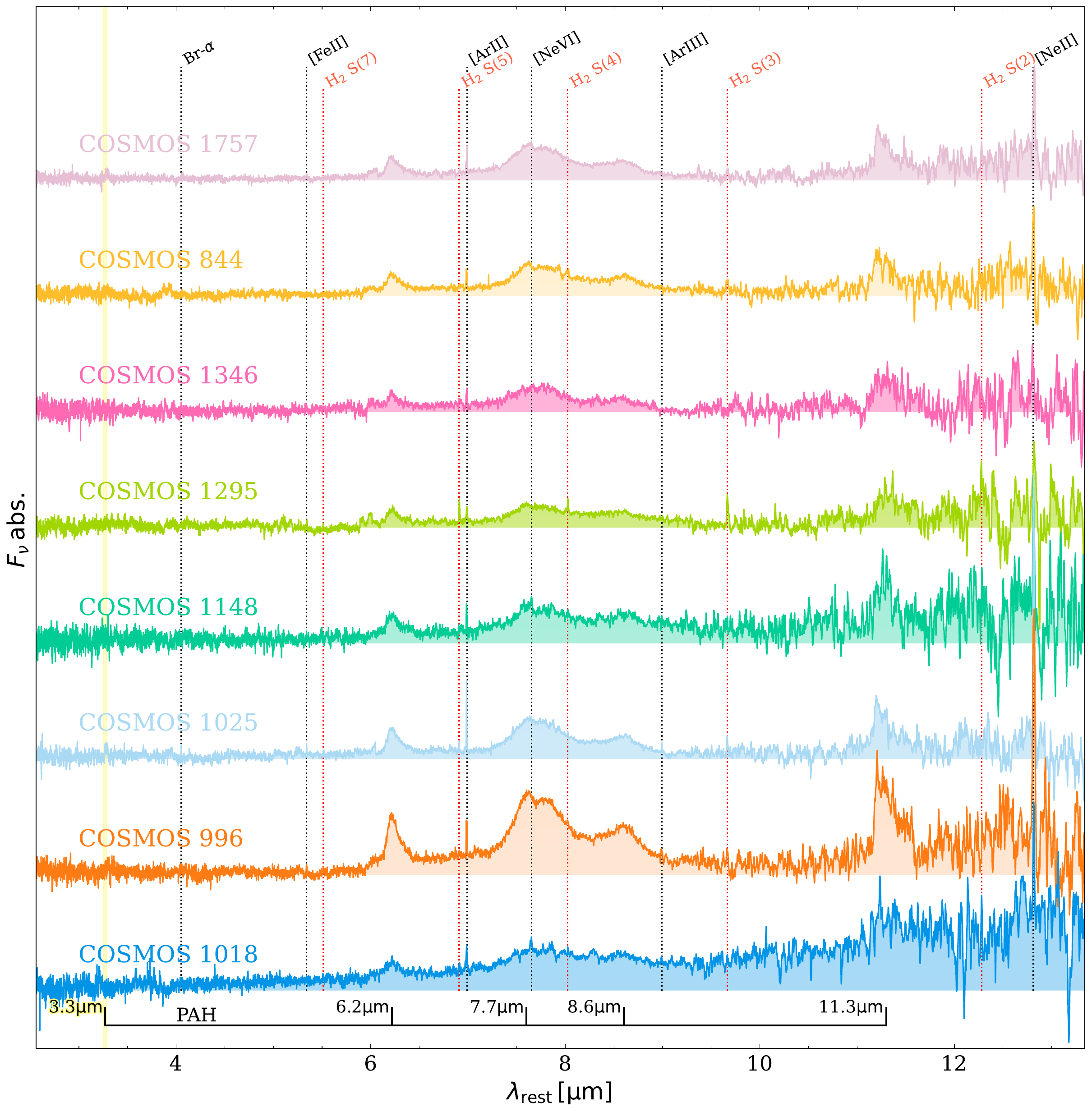}
    \caption{Collage of integrated JWST/MIRI MRS spectra for our sample extracted. Several fine-structure lines, H$_2$ rotational transitions and PAH bands are indicated.}
    \label{fig:spc_present_app}
\end{figure*}

\begin{figure*}
    \centering
    \includegraphics[width=\textwidth,clip]{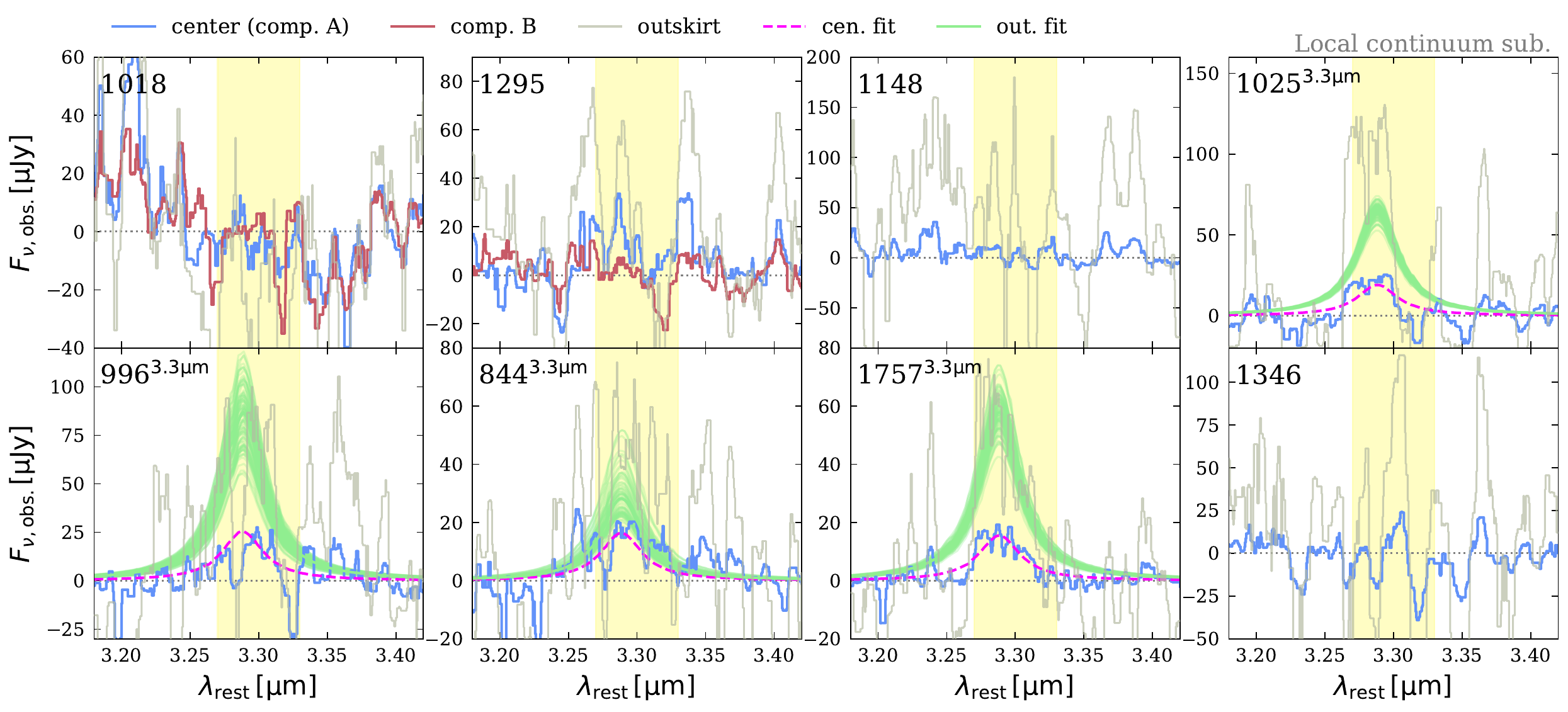}
    \caption{Zoom-in on PAH$3.3\,{\rm \mu m}$ with local continuum subtracted. The PAH$3.3\,{\rm \mu m}$ band emission is fit separately from the other PAH bands using \texttt{CAFE}. Galaxies with $3.3\,{\rm \mu m}$ detections are marked.}
    \label{fig:pah33_fit}
\end{figure*}

We show the integrated total spectra for our sample in Fig. \ref{fig:spc_present_app}. The spectrum of each galaxy equals to the summations of the spectra from the center (or both merger components) and the outskirt regions (Fig. \ref{fig:spc_present}). We show the fit of PAH3.3$\,\mu$m in Fig. \ref{fig:pah33_fit}.

\begin{figure*}
    \centering
    \includegraphics[width=\textwidth,clip]{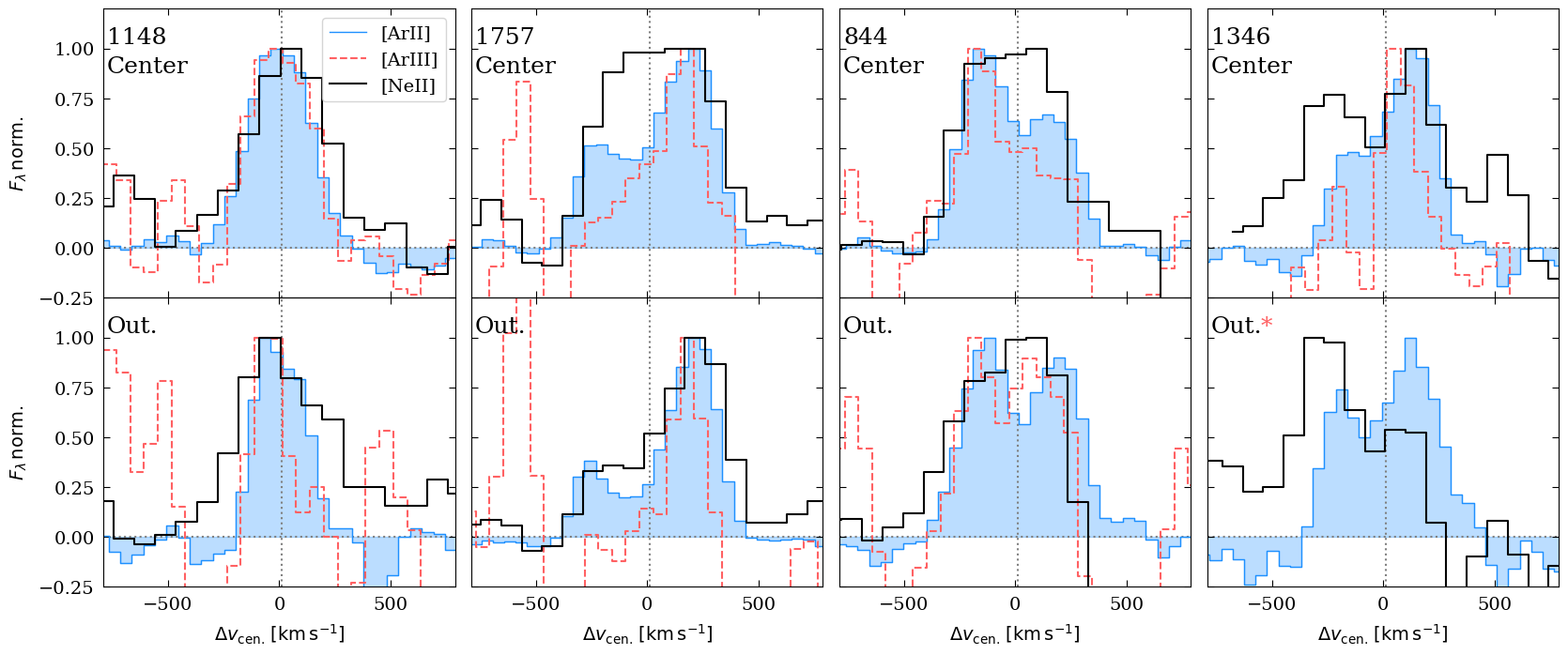}
    \caption{Similar to Fig. \ref{fig:spc_line_nonmer} but for the remaining non-merger sample. The [\ion{Ar}{iii}] from the outskirt of COSMOS 1346 is a non-detection.}
\label{fig:spc_li_app}
\end{figure*}

In Fig. \ref{fig:spc_li_app}, we present the ionization lines for COSMOS 1148, 1757, 884, and 1346. Similar to Fig. \ref{fig:spc_line_mer}, we show [\ion{Ar}{ii}], [\ion{A}{iii}], and [\ion{Ne}{ii}].

\begin{figure*}
    \centering
    \includegraphics[width=0.8\textwidth,clip]{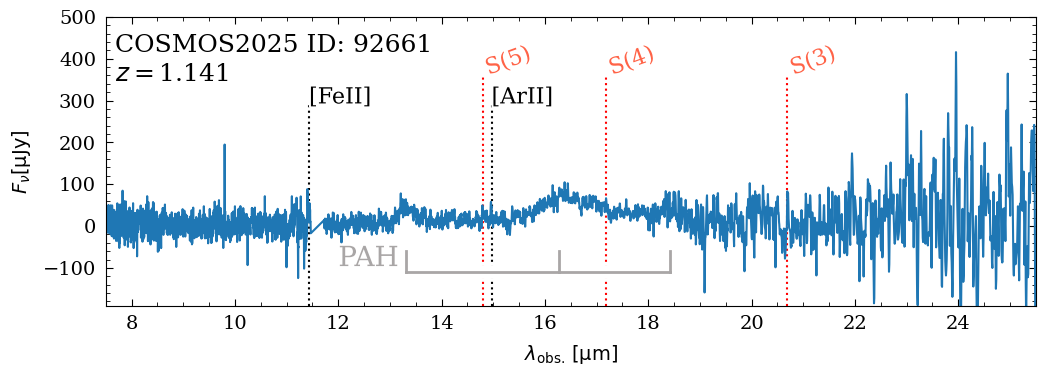}
    \caption{JWST/MIRI MRS integrated spectrum of the serendipitous galaxy detection near COSMOS 1018 at $z=1.141$ (identified via the \ion{Ar}{ii} fine-structure line. The ID of this galaxy in the catalog COSMOS2025 is 92661 \citep[][]{Shuntov_2025}.}
\label{fig:near_c18}
\end{figure*}

The integrated MIRI MRS spectrum of the $z=1.141$ galaxy near COSMOS 1018 is presented in Fig. \ref{fig:near_c18}. We mark emission lines with detections. The redshift is based on a single-Gaussian fitting of its [\ion{Ar}{ii}]. The extraction aperture is shown in Fig. \ref{fig:nircam_rgb}. 

\section{Fitting Results of Mid-IR Emission Lines}\label{app:line_fit}
We present emission line fitting results in Table \ref{tab:line_fit}. For non-detections, we report $3\sigma$ upper limits.

\newpage
\begin{sidewaystable}
\centering
\scriptsize
\setlength{\tabcolsep}{3pt}
\caption{Observed emission line fluxes in $10^{-17}\,{\rm erg\,s^{-1}\,cm^{-2}}$.}
\label{tab:line_fit}
\begin{tabular}{lccccccccccc} 
\hline
\textbf{COSMOS \#} & Br$\alpha$ & [\ion{Fe}{ii}] & [\ion{Ar}{ii}]  & [\ion{Ne}{vi}]  & [\ion{Ar}{iii}] & [\ion{Ne}{ii}]  & $\rm H_{2}(S2)$ & $\rm H_{2}(S3)$ & $\rm H_{2}(S4)$ & $\rm H_{2}(S5)$ & $\rm H_{2}(S7)$ \\
\hline
\textbf{1018} &  &  &  &  &  &  &  &  &  &  &  \\
\cline{1-1}
\quad Comp. A & $<2.97$ & $0.63^{+0.16}_{-0.11}$ & $4.00^{+0.08}_{-0.07}$ & $3.85^{+0.15}_{-0.20}$ & $1.11^{+0.05}_{-0.04}$ & $7.82^{+1.19}_{-1.25}$ & $1.96^{+0.91}_{-0.71}$ & $1.88^{+0.36}_{-0.27}$ & $0.74^{+0.02}_{-0.02}$ & $1.76^{+0.04}_{-0.05}$ & $1.54^{+0.22}_{-0.14}$ \\
\quad Comp. B & $<2.76$ & $0.82^{+0.10}_{-0.09}$ & $3.48^{+0.11}_{-0.12}$ & $1.55^{+0.09}_{-0.08}$ & $<0.61$ & $5.01^{+1.30}_{-1.05}$ & $3.97^{+1.00}_{-0.87}$ & $<1.13$ & $0.30^{+0.02}_{-0.01}$ & $0.93^{+0.12}_{-0.09}$ & $0.73^{+0.09}_{-0.10}$ \\
\quad Outskirt & $<8.86$ & $<3.67$ & $6.52^{+0.29}_{-0.24}$ & $4.66^{+0.09}_{-0.13}$ & $1.38^{+0.07}_{-0.07}$ & $12.13^{+3.43}_{-1.45}$ & $3.69^{+1.79}_{-3.69}$ & $<2.34$ & $1.16^{+0.04}_{-0.03}$ & $1.96^{+0.12}_{-0.11}$ & $1.54^{+0.49}_{-0.44}$ \\

\textbf{1295} &  &  &  &  &  &  &  &  &  &  &  \\
\cline{1-1}
\quad Comp. A & $<1.66$ & $1.58^{+0.21}_{-0.19}$ & $3.23^{+0.02}_{-0.03}$ & $<0.41$ & $<0.36$ & $10.84^{+1.74}_{-1.45}$ & $3.51^{+0.90}_{-0.71}$ & $4.00^{+0.06}_{-0.07}$ & $1.90^{+0.02}_{-0.02}$ & $3.77^{+0.02}_{-0.02}$ & $2.11^{+0.07}_{-0.10}$ \\
\quad Comp. B & $<1.74$ & $1.29^{+0.17}_{-0.33}$ & $2.17^{+0.10}_{-0.38}$ & $<0.33$ & $0.32^{+0.01}_{-0.01}$ & $11.15^{+2.03}_{-1.80}$ & $<2.81$ & $3.12^{+0.10}_{-0.10}$ & $1.18^{+0.03}_{-0.03}$ & $2.43^{+0.02}_{-0.03}$ & $1.16^{+0.11}_{-0.08}$ \\
\quad Outskirt & $<4.61$ & $3.23^{+0.36}_{-0.32}$ & $5.72^{+0.12}_{-0.16}$ & $<0.77$ & $0.65^{+0.04}_{-0.04}$ & $28.73^{+2.94}_{-2.88}$ & $7.65^{+2.19}_{-0.60}$ & $10.53^{+0.22}_{-0.31}$ & $3.70^{+0.08}_{-0.06}$ & $6.56^{+0.06}_{-0.07}$ & $2.66^{+0.20}_{-0.16}$ \\

\textbf{1148} &  &  &  &  &  &  &  &  &  &  &  \\
\cline{1-1}
\quad Center & $<2.84$ & $0.82^{+0.10}_{-0.08}$ & $2.96^{+0.05}_{-0.05}$ & $2.90^{+0.05}_{-0.05}$ & $0.64^{+0.03}_{-0.03}$ & $11.06^{+2.18}_{-2.04}$ & $<5.28$ & $<1.53$ & $0.69^{+0.02}_{-0.02}$ & $1.37^{+0.05}_{-0.05}$ & $0.83^{+0.12}_{-0.10}$ \\
\quad Outskirt & $<10.64$ & $1.56^{+0.91}_{-0.94}$ & $7.52^{+0.09}_{-0.11}$ & $6.89^{+0.16}_{-0.14}$ & $0.89^{+0.08}_{-0.07}$ & $36.62^{+14.48}_{-6.85}$ & $<24.67$ & $<6.34$ & $<2.67$ & $<3.75$ & $<5.71$ \\

\textbf{1025} &  &  &  &  &  &  &  &  &  &  &  \\
\cline{1-1}
\quad Center & $1.11^{+0.10}_{-0.10}$ & $1.04^{+0.03}_{-0.03}$ & $4.41^{+0.01}_{-0.01}$ & $<0.66$ & $0.50^{+0.01}_{-0.01}$ & $13.36^{+0.35}_{-0.26}$ & $<2.60$ & $1.34^{+0.07}_{-0.06}$ & $0.24^{+0.01}_{-0.01}$ & $0.54^{+0.01}_{-0.01}$ & $<0.55$ \\
\quad Outskirt & $3.45^{+0.31}_{-0.27}$ & $2.42^{+0.09}_{-0.08}$ & $13.10^{+0.03}_{-0.03}$ & $<1.85$ & $1.67^{+0.02}_{-0.02}$ & $36.53^{+0.62}_{-0.83}$ & $<4.65$ & $3.84^{+0.09}_{-0.11}$ & $0.55^{+0.01}_{-0.01}$ & $1.75^{+0.02}_{-0.02}$ & $<2.78$ \\

\textbf{996} &  &  &  &  &  &  &  &  &  &  &  \\
\cline{1-1}
\quad Center & $<3.14$ & $1.99^{+0.14}_{-0.12}$ & $6.48^{+0.07}_{-0.05}$ & $<0.43$ & $1.04^{+0.02}_{-0.03}$ & $27.01^{+0.92}_{-1.13}$ & $<3.66$ & $2.07^{+0.19}_{-0.12}$ & $0.80^{+0.03}_{-0.02}$ & $0.78^{+0.04}_{-0.04}$ & $<1.65$ \\
\quad Outskirt & $<10.06$ & $7.15^{+0.45}_{-0.44}$ & $15.85^{+0.15}_{-0.10}$ & $<1.72$ & $3.24^{+0.07}_{-0.08}$ & $81.35^{+4.66}_{-5.72}$ & $<16.06$ & $9.45^{+0.57}_{-0.60}$ & $2.14^{+0.06}_{-0.05}$ & $<1.96$ & $<3.73$ \\

\textbf{844} &  &  &  &  &  &  &  &  &  &  &  \\
\cline{1-1}
\quad Center  & $<2.91$ & $0.82^{+0.08}_{-0.08}$ & $3.36^{+0.11}_{-0.08}$ & $<0.37$ & $0.28^{+0.02}_{-0.02}$ & $8.42^{+0.71}_{-0.81}$ & $<2.59$ & $1.38^{+0.12}_{-0.07}$ & $<0.29$ & $<0.38$ & $<0.43$ \\
\quad Outskirt & $<7.41$ & $1.51^{+0.49}_{-0.45}$ & $8.89^{+0.12}_{-0.17}$ & $<2.16$ & $1.21^{+0.04}_{-0.05}$ & $18.89^{+3.61}_{-3.00}$ & $<8.78$ & $6.47^{+0.27}_{-0.29}$ & $<1.70$ & $<1.81$ & $<2.01$ \\

\textbf{1757} &  &  &  &  &  &  &  &  &  &  &  \\
\cline{1-1}
\quad Center  & $<1.42$ & $0.79^{+0.09}_{-0.08}$ & $3.46^{+0.02}_{-0.02}$ & $<0.25$ & $0.11^{+0.01}_{-0.01}$ & $8.99^{+1.39}_{-0.71}$ & $<2.61$ & $1.21^{+0.10}_{-0.11}$ & $<0.23$ & $0.62^{+0.02}_{-0.03}$ & $0.35^{+0.04}_{-0.04}$ \\
\quad Outskirt & $<8.95$ & $2.25^{+0.21}_{-0.20}$ & $9.37^{+0.06}_{-0.06}$ & $<1.43$ & $0.26^{+0.02}_{-0.03}$ & $24.98^{+1.39}_{-1.93}$ & $3.62^{+1.02}_{-0.76}$ & $4.22^{+0.25}_{-0.18}$ & $<0.68$ & $1.73^{+0.06}_{-0.07}$ & $<1.72$ \\

\textbf{1346} &  &  &  &  &  &  &  &  &  &  &  \\
\cline{1-1}
\quad Center & $1.72^{+0.62}_{-0.32}$ & $1.04^{+0.10}_{-0.16}$ & $2.16^{+0.03}_{-0.02}$ & $<0.34$ & $0.16^{+0.01}_{-0.01}$ & $5.14^{+0.49}_{-1.08}$ & $<2.13$ & $<0.99$ & $<0.36$ & $<1.03$ & $<0.53$ \\
\quad Outskirt & $10.72^{+2.85}_{-2.70}$ & $<4.34$ & $6.91^{+0.10}_{-0.10}$ & $<2.25$ & $<0.82$ & $18.16^{+3.54}_{-2.89}$ & $<7.67$ & $<2.74$ & $<1.44$ & $<4.17$ & $<2.14$ \\

\hline
\end{tabular}
\end{sidewaystable}
\newpage

\section{Low mid-IR Dust Attenuation of our sample}\label{app:atten_check}
\begin{figure*}
    \centering
    \includegraphics[width=\textwidth,clip]{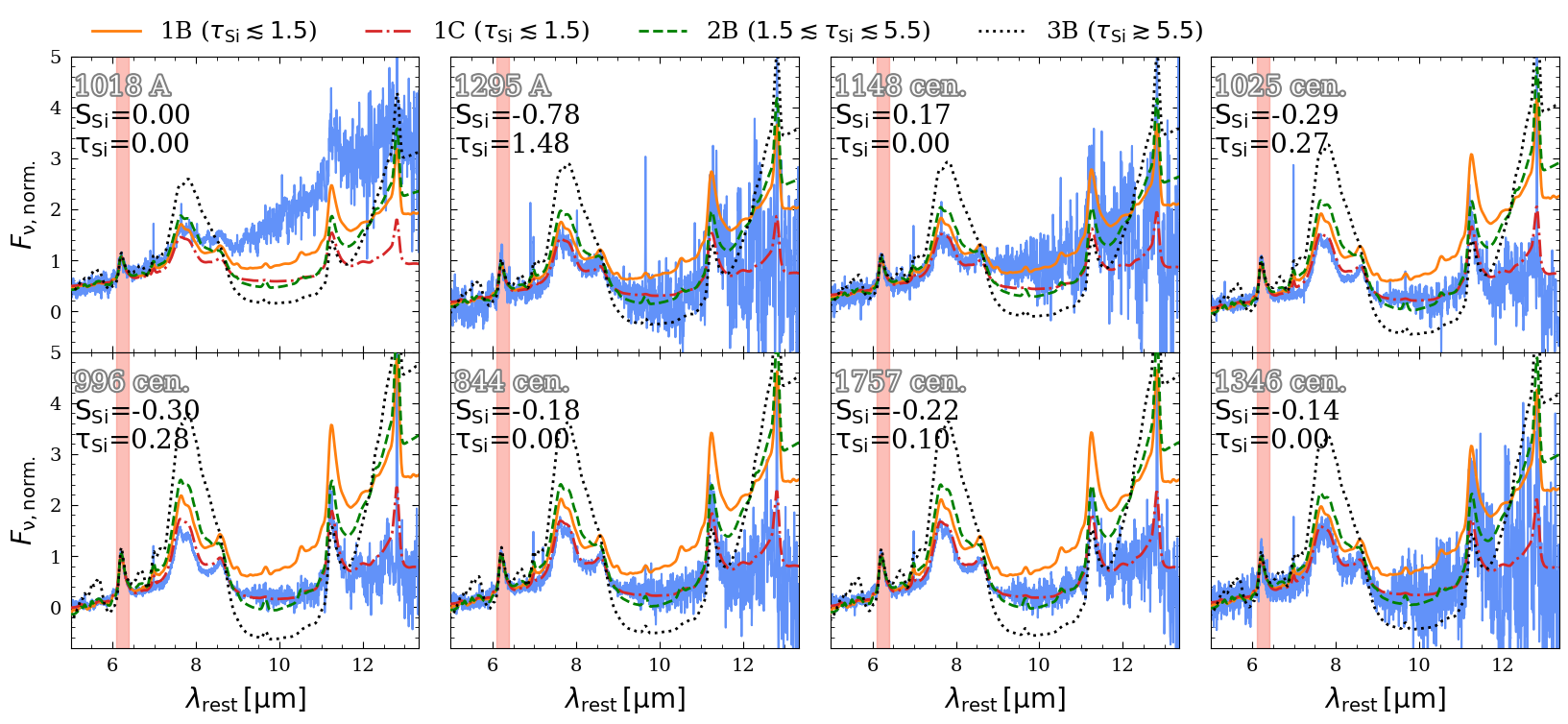}
    \caption{
    Comparison of JWST/MIRI MRS spectra of our targets to four different empirical templates with varying $\tau_{\rm Si}$ dust attenuation from \citet{Lai_2020}. The templates are normalized to the integrated PAH$6.2\,{\rm \mu m}$ flux and matched continuum at $6.2\rm \mu m$, i.e., same equivalent width of PAH$6.2\,{\rm \mu m}$. From this comparison we conclude that the galaxies in our sample have likely a low silicate dust attenuation on the order of $\tau_{\rm Si} < 1.5$. Note that COSMOS~1018 is an AGN with rising mid-IR host dust continuum, which complicates the comparison to the empirical templates.}
\label{fig:atten_check}
\end{figure*}

As summarized in Sect. \ref{subsec:dis_dust_att}, we used two methods to validate the assumption of $\tau_{\rm Si}$ when fitting with \texttt{CAFE}. First, we quantify the silicate strength $S_{\rm Si}=\ln(f_{\rm 9.7\mu m,obs}/f_{\rm 9.7\mu m,cont})$. The $f_{\rm 9.7\mu m,cont}$ is estimated with with polynomial interpolation using wavelength regions less affected by absorption following \citet{Spoon_2007}. Specifically, we fit a second order polynomial for the continuum using median flux from four 0.5$\,\rm\mu$m wide wavelength ranges around rest frame 2.8, 4.7, 5.4, and 13.4$\,\rm\mu$m. The $f_{\rm 9.7\mu m,cont}$ is interpolated as the median flux at 0.3$\,\rm\mu$m around 9.7$\,\rm\mu$m. The observed $f_{\rm 9.7\mu m,obs}$ is calculated in the same range from observed spectra. We then convert the $S_{\rm Si}$ to $\tau_{\rm Si}$ using $\tau_{\rm  Si}=-2.48\times S_{\rm Si}-0.45$ providing $\tau_{\rm  Si}\geq0$ \citep[][]{Lai_2024}. We show resulted $S_{\rm Si}$ and $\tau_{\rm Si}$ in Fig. \ref{fig:atten_check} for the center regions (and brighter merger component A) as the outskirt is noisy and expected to be lower in attenuation if any. Most of our targets have $\tau_{\rm Si}<0.3$ with only COSMOS 1295A at $\tau_{\rm Si}\sim1.5$. We note that the here estimated $\tau_{\rm Si}$ to the zeroth order should only be treated as an approximation which depends on the degree of the polynomial and wavelength range longer than the silicate absorption. For example, if using a 3rd polynomial, COSMOS 1295A will have a $\tau_{\rm Si}\sim0.3$. Due to low sensitivity and MIRI thermal background, the only available wavelength range around 13.4$\,\rm\mu$m also show an over estimation of the continuum flux which biasing the $\tau_{\rm Si}$ towards higher values. 

Independently, we compare our spectra against the \citet[][]{Lai_2020} templates to qualitatively assess the obscuration of our sources. Specifically, the templates are classified into nine categories based on equivalent width of PAH$6.2\,\mu$m ($\rm EW_{PAH6.2\,\mu m}$) and $\rm S_{Si}$ \citep[][]{Spoon_2007}. The number 1, 2, and 3 indicate the three sub-categories of $\rm S_{Si}$ or $\rm \tau_{Si}$ with  $\rm \tau_{Si}\lesssim1.5$, $\rm \lesssim1.5\tau_{Si}\lesssim5.5$, and $\rm \tau_{Si}\gtrsim5.5$, respectively. The letters A, B, and C mark the $\rm EW_{PAH6.2\,\mu m}$ in ascending order. Based on visual check, we pre-exclude continuum-dominated templates (i.e., As with no obvious PAH features). We normalize the templates of 1B, 1C, 2B, and 3B to the same $\rm EW_{PAH6.2\,\mu m}$ as our targets and show the comparison in Fig. \ref{fig:atten_check}. We find qualitatively that all but COSMOS 1018 match to 1B or 1C. It is obvious that the continuum of COSMOS 1018, dominated by hot AGN dust (Sect. \ref{subsec:res_AGN}), has no sign of silicate absorption as indicated by the $\tau_{\rm Si}$ quantified above. Hence, this comparison implies that our sample has at least $\rm \tau_{Si}\lesssim1.5$.

While we cannot directly constrain the attenuation, the results from the two methods suggest our assumption of low mid-IR dust attenuation is reasonable. In Sect. \ref{subsec:dis_dust_att}, we further quantify the potential impact if $\tau_{\rm Si}>0$.



\bibliography{references}{}
\bibliographystyle{aasjournalv7}



\end{document}